\documentclass[3p]{elsarticle}
\pdfoutput=1
\usepackage[utf8]{inputenc}
\usepackage{graphicx}
\usepackage{subcaption}  
\usepackage{hyperref, amsmath, amsfonts, natbib, bm, floatrow, multirow, wrapfig}
\usepackage[dvipsnames]{xcolor}
\usepackage{ulem} 
\usepackage{algorithm}
\usepackage{algpseudocode}

\usepackage{caption}
\usepackage{subcaption}
\usepackage{float}
\usepackage{booktabs}

\graphicspath{{Figures/}}

\journal{Journal name}

\begin{document}

\begin{frontmatter}

\title{Modelisation of chaotic systems with a latent Stochastic Differential Equation}

\author[inst1,inst2]{Ismaël Zighed\corref{cor1}}
\ead{ismael.zighed@sorbonne-universite.fr}

\author[inst2]{Nicolas Thome}
\ead{nicolas.thome@sorbonne-universite.fr}

\author[inst2]{Patrick Gallinari}
\ead{patrick.gallinari@sorbonne-universite.fr}

\author[inst3]{Taraneh Sayadi}
\ead{taraneh.sayadi@lecnam.net}

\cortext[cor1]{Corresponding authors}

\affiliation[inst1]{organization={Institut Jean le Rond d'Alembert, Sorbonne Université},
            country={France}}

\affiliation[inst2]{organization={ISIR, Sorbonne Université},
            country={France}}

\affiliation[inst3]{organization={M2N, Conservatoire National des Arts et Métiers},
            country={France}}

\begin{abstract}
Stochastic Differential Equations (SDEs) have become a cornerstone of scientific machine learning, though they are predominantly utilized as algorithmic tools for uncertainty quantification or distribution matching. In contrast, leveraging SDEs fundamentally to model macroscopic, nonlinear physics as stochastic processes remains largely unexplored. This work introduces a probabilistic, non-intrusive reduced-order model (ROM) for chaotic dynamical systems.

We argue that projecting high-dimensional nonlinear dynamics onto a low-dimensional manifold introduces irreducible uncertainty, compounded by the chaotic attractors and multi-admissible futures inherent to turbulent flows. Consequently, a chaotic system governed by a partial differential equation can be effectively modeled by an SDE in a suitable latent space. To this end, a nonlinear autoencoder is employed to map the flow field into a low-dimensional representation, within which the temporal evolution is explicitly governed by an SDE. The predictable component of the dynamics is captured by a learned residual drift term, while state-dependent stochasticity is absorbed by a diffusion term.

We demonstrate that this probabilistic framework successfully propagates highly nonlinear states, offering a robust alternative to traditional deterministic methodologies for chaotic regimes. Ultimately, our model generates new chaotic flow trajectories that remain locally and globally consistent with the true transition kernel learned from Direct Numerical Simulation (DNS) data. Even though these generated trajectories are unique and distinct from the training set, they preserve the underlying statistics and manifolds, validating the strong generative performance and robustness of our methodology. Our code,  one experimental dataset along with its associated pre-trained model are publicly available on \href{https://github.com/IsmaelZig-SU/DriftNDiffuse}{Github}.
\end{abstract}


\end{frontmatter}

\section{Introduction}

This work explores reduced-order models (ROMs) for dynamical systems through a probabilistic lens. Traditional methodologies often rely on solving equations in a reduced space \citep{POD-Galerkin, CD-ROM, quantized-ROM}, searching for a closed Koopman space \citep{Koopman, DMD, MZAE}, or constructing deterministic, non-linear reduced dynamical systems from data \citep{UPdROM, ESN, LSTM, Transformer, NODE}. In contrast, we ask: can one condition the future state of a reduced system entirely on its past? Furthermore, can we learn this underlying distribution and sample from it to propagate a highly non-linear system within a low-dimensional latent space? This work demonstrates that, through relatively simple modeling choices, a probabilistic framework can challenge traditional deterministic approaches by capturing not only model uncertainties but also the intrinsic stochastic nature of the system. This motivates the adoption of a strategy in which the dynamics are modeled stochastically, through the use of stochastic differential equations (SDEs).

Consider a high-dimensional, highly nonlinear dynamical system. Projecting this system onto a reduced manifold is the cornerstone of reduced-order modelling, essential for maintaining fast and computationally cheap training and inference. However, these manifolds often exhibit complex topologies \citep{SSM}, multiple, chaotic or strange attractors \citep{chaotic-SSM}. Even within this simplified, reduced space, a system may exhibit multi-admissible futures, extreme sensitivity to initial conditions, and a sense of stochasticity stemming inherently from its chaotic nature and from information lost in the projection's orthogonal subspace \citep{Mori, Zwanzig}. Consequently, these systems, and their reduced counterparts, harbour an irreducible amount of uncertainty.
To illustrate, imagine a ball balanced perfectly atop a symmetrical mountain: will it roll down the left side or the right? Traditional techniques attempt to refine the system's resolution to capture the most intricate initial conditions, aiming to determine whether the ball leans microscopically to one side. Instead, our framework accepts that the ball has an equal probability of falling either way. Once this initial "decision" is made, subsequent states can be computed within a Bayesian framework. This work applies a similar conceptual shift to chaotic fluid flows, where randomness and multi-admissible futures are framed as a chain of probabilistic transitions. These states possess a predictable component inherited from the system's deterministic physics, alongside an inherent stochastic component \citep{SDE}. Rather than attempting to close the system deterministically, we embrace this stochasticity, framing the state of our reduced system as the flow of a probability distribution in a latent space. Specifically, the reduced system is modelled via a Stochastic Differential Equation (SDE), where the predictable dynamics are governed by a drift term and the stochastic variations are captured by a diffusion term. This diffusion term can \textit{de-facto} be learnt, as its magnitude and entropy, are state (or time)-dependent, maximizing when the ball is balanced precariously at the peak, and minimizing as it accelerates down the steep slope. The aim then, is to learn the transition kernel. Stochastic Differential Equations (SDEs) have become relevant in several machine learning paradigms. For instance, they are widely used for uncertainty quantification in autoregressive rollouts, where accumulating error is modelled via a diffusion term \citep{Tornado}, and can even separate aleatoric and epistemic uncertainties \citep{SDE-net}, treating the model's transformations itself as an SDE. Algorithmically, SDEs are also heavily leveraged to steer an initial distribution toward a target data distribution, as seen in diffusion models \citep{Diffusion} and their continuous-time counterpart \cite{score_matching}. While these approaches use SDEs instrumentally to improve algorithmic performance, we employ SDEs fundamentally to capture the intrinsic physical ``randomness" of the system within its latent representation. This paper is organized as follows:  Section \ref{sec:background} skims the existing research this work builds upon. Section  \ref{sec:dynROM} establishes our methodology. Section \ref{sec:experiments} describes the chaotic problems used as our benchmark. Section \ref{sec:results} presents the experimental results and a detailed discussion. Section \ref{sec:conclusion} concludes the paper and outlines potential avenues for future work. \\

\section{Background}
\label{sec:background}

The foundation of this study builds upon the existing works in latent Stochastic Differential Equations, also referred to as SDE matching, which formulates variational inference over the infinite-dimensional path-space of continuous-time dynamical systems. Applying variational inference within such spaces presents a set of mathematical and algorithmic challenges. From a mathematical standpoint, \citep{AdaptivePathIntegral} derives the loss function for such latent SDE emulators.

Let $Z = \{z_t\}_{t \in [0, T]}$ be a continuous latent trajectory in $\mathbb{R}^d$ over $[0, T]$, and let $\Phi = \{\varphi_1, \dots, \varphi_K\}$ be a sequence of $K$ high-dimensional observations, with $\varphi_k \in \mathbb{R}^{d_x}$ ($d_x \gg d$) recorded at time $t_k$ and generated by the latent state $z_{t_k}$.

An SDE matching model, parameterized by $\theta$, aims to maximize the marginal log-likelihood of the data. The problem is that this quantity requires marginalizing over every possible latent trajectory $Z$, making it analytically intractable:
\begin{align}
    \theta^* &= \arg \max_\theta \log p_\theta (\Phi), \label{eq:LL_SDE_matching} \\
    p_\theta (\Phi) &= \int p_\theta (\Phi \mid Z) \, d\mathbb{P}_\theta(Z). \label{eq:LL_SDE_matching_2}
\end{align}
Where $\mathbb{P}_\theta$ is the prior, infinite-dimensional path measure. Assuming the observations are conditionally independent given the trajectory, and that the prior's dynamics are governed by an Itô SDE, we have:
\begin{align}
    p_\theta(\Phi \mid Z) &= \prod_{k=1}^K p_\theta(\varphi_k \mid z_{t_k}), \\
    dz_t &= f_\theta(z_t, t)\,dt + g_\theta(z_t, t)\,dW_t, \qquad z_0 \sim p_\theta(z_0). \label{eq:SDE_background}
\end{align}
Because $\mathbb{P}_\theta$ has no closed-form transition density, Equation~\ref{eq:LL_SDE_matching_2} simply cannot be evaluated directly, and naive Monte Carlo sampling is hopeless as the odds of a random prior path landing close to the data are minimal. \citep{AdaptivePathIntegral} works around this by introducing an approximate posterior, parametrized by $\xi$, which defines a new path measure $\mathbb{Q}_\xi$ through a surrogate SDE:
\begin{equation}
    dz_t = h_\xi(z_t, t, \Phi)\,dt + g_\theta(z_t, t)\,dW_t, \qquad z_0 \sim q_\xi(z_0 \mid \Phi).
    \label{eq:surrogate_SDE}
\end{equation}
Note that the diffusion term $g_\theta(z_t, t)$ matches the prior's exactly. This stems from a modelling constraint which guarantees that $\mathbb{Q}_\xi$ stays absolutely continuous with respect to $\mathbb{P}_\theta$ ($\mathbb{Q}_\xi \ll \mathbb{P}_\theta$), which in turn makes the Radon–Nikodym derivative $d\mathbb{P}_\theta / d\mathbb{Q}_\xi$ well-defined.

With a change of measure and Jensen's inequality, we get the Evidence Lower Bound (ELBO), which then splits cleanly into a reconstruction term and a KL term:
\begin{align}
    \log p_\theta(\Phi) &= \log \mathbb{E}_{\mathbb{Q}_\xi} \left[ p_\theta(\Phi \mid Z) \frac{d\mathbb{P}_\theta}{d\mathbb{Q}_\xi}(Z) \right] \geq \mathbb{E}_{\mathbb{Q}_\xi} \left[ \log p_\theta(\Phi \mid Z) + \log \frac{d\mathbb{P}_\theta}{d\mathbb{Q}_\xi}(Z) \right], \\
    \mathcal{L}_{\text{ELBO}} &= \mathbb{E}_{\mathbb{Q}_\xi} \left[ \sum_{k=1}^K \log p_\theta(\varphi_k \mid z_{t_k}) \right] - D_{\text{KL}}(\mathbb{Q}_\xi \parallel \mathbb{P}_\theta).
\end{align}
The first term is a perfectly tractable reconstruction loss, evaluated at the discrete observation times. It is conceptually a decoder loss. The second is a Kullback-Leibler divergence expressed in terms of the Radon–Nikodym derivative. The KL divergence between two infinite-dimensional path measures is infinite whenever the diffusion terms are not equal, due to the absolute continuity condition. This motivates the diffusion term used in ~\ref{eq:surrogate_SDE}. This term is therefore finite yet intractable in its current form. The Girsanov theorem turns this KL term into a tractable integral (see Appendix~\ref{appendix:Girsanov_SOC} for the derivation):
\begin{equation}
    D_{\text{KL}}(\mathbb{Q}_\xi \parallel \mathbb{P}_\theta) = D_{\text{KL}}(q_\xi(z_0 \mid \Phi) \parallel p_\theta(z_0)) + \frac{1}{2} \mathbb{E}_{\mathbb{Q}_\xi} \left[ \int_0^T \|r_{\theta, \xi}(z_t, t)\|_2^2 \, dt \right].
\end{equation}
With $r_{\theta, \xi}(z_t, t) = g_\theta^{-1}(z_t, t) \big( h_\xi(z_t, t, \Phi) - f_\theta(z_t, t) \big)$ the drift discrepancy. The loss function for SDE matching, written as the negative of the log-likelihood, then writes : 
\begin{equation}
\label{eq:loss_SDEmatching}
    \mathcal{L}_{\text{loss}} = 
    \underbrace{-\mathbb{E}_{\mathbb{Q}_\xi} \left[ \sum_{k=1}^K \log p_\theta(\varphi_k \mid z_{t_k}) \right]}_{\mathcal{L}_{\text{recon}}}
    \;+\;
    \underbrace{D_{\text{KL}}(q_\xi(z_0 \mid \Phi) \parallel p_\theta(z_0))}_{\mathcal{L}_{\text{prior}}}
    \;+\;
    \underbrace{\frac{1}{2} \mathbb{E}_{\mathbb{Q}_\xi} \left[ \int_0^T \|r_{\theta, \xi}(z_t, t)\|_2^2 \, dt \right]}_{\mathcal{L}_{\text{diff}}}.
\end{equation}
where $\mathcal{L}_{\text{recon}}$ handles reconstruction error at the observation times, $\mathcal{L}_{\text{prior}}$ regularizes the inferred initial condition, and $\mathcal{L}_{\text{diff}}$ penalizes any drift between the learned surrogate dynamics and the true prior dynamics over the whole interval.

This objective is elegant on paper, but expensive in practice. Since $\mathcal{L}_{\text{diff}}$ is a time integral, computing it at each training step means simulating the surrogate SDE forward over the full interval $[0, T]$ with a numerical solver. Backpropagating through this continuous-time integral generally requires adjoint sensitivity methods, which avoid the memory blowup of standard backpropagation through time, but at the cost of solving a backward-in-time adjoint SDE, introducing computational and numerical bottlenecks.

Recent work sidesteps this computational bottleneck entirely through simulation-free training. \citep{CourseNair} show that if the surrogate posterior is restricted to a linear (Markov Gaussian) SDE, its marginal distribution $q_\xi(z_t \mid t)$ at any time $t$ is available in closed form, with the mean and covariance evolving according to simple ODEs. This makes it possible to evaluate the drift residual $r_{\theta, \xi}$ analytically at arbitrary time points, so the model can be trained by sampling directly from these Gaussian marginals instead of numerically integrating the SDE, eliminating the forward solver, and the adjoint backward pass, from the training loop altogether. \citep{DDStoROM} then extend this work to parametric dependency and forcing to dynamical systems. An alternative simulation-free strategy is proposed by \citep{SDEMatching}, drawing on ideas from continuous-time diffusion and score/flow matching \citep{score_matching, FM}. Rather than restricting the posterior to a linear SDE, \citep{SDEMatching} define a generative flow $z_t = F_\xi(\epsilon, t, \Phi)$, $\epsilon \sim \mathcal{N}(0, I)$, whose time-derivative along fixed noise $\epsilon$ yields a velocity field $v_\xi$ generating the marginals $q_\xi(z_t \mid \Phi)$:
\begin{equation}
    \frac{d}{dt} z_t = v_\xi(z_t, t, \Phi).
\end{equation}
Since $F_\xi(\cdot, t, \Phi)$ is invertible in $\epsilon$, the change-of-variables formula gives the marginal log-density in closed form:
\begin{equation}
    \log q_\xi(z_t \mid \Phi) = \log p(\epsilon) - \log \left| \det \frac{\partial F_\xi(\epsilon, t, \Phi)}{\partial \epsilon} \right|, \qquad \epsilon = F_\xi^{-1}(z_t, t, \Phi).
\end{equation}
Using the equivalence between an SDE's marginals and those of its associated probability-flow ODE, the posterior drift $h_\xi$ consistent with both $v_\xi$ and the shared diffusion term $g_\theta$ can then be recovered in closed form:
\begin{equation}
    h_\xi(z_t, t, \Phi) = v_\xi(z_t, t, \Phi) + \frac{1}{2} g_\theta(z_t, t) g_\theta(z_t, t)^T \nabla_{z_t} \log q_\xi(z_t \mid \Phi).
\end{equation}
This enables closed-form evaluation, at any arbitrary $t$, of the drift-matching loss:
\begin{equation}
    \mathcal{L}_{\text{diff}} = \mathbb{E}_{t, \epsilon} \frac{1}{2} \left\| g_\theta^{-1}(z_t, t) (  h_\xi(z_t, t, \Phi) - f_\theta(z_t, t) ) \right\|_2^2.
\end{equation}

Ultimately, all these methods aim to bypass the intractable integral term $\mathcal{L}_{\text{diff}}$ in Equation~\ref{eq:loss_SDEmatching}. In contrast, we pursue a considerably simpler formulation: rather than approximating the continuous-time SDE posterior, we take the discrete-time, deterministic-autoencoder limit of this framework. This alleviates the need for forward SDE pass and adjoint sensitivity back-propagation, while considerably simplifying the model and accelerating the training. We discretize the dynamics using an Euler-Maruyama scheme with a fixed time increment $\Delta t = 1$.  We also fit the SDE posterior directly to the latent data (as a consequence of the deterministic autoencoder), hence we do not distinguish a prior SDE and measure $\mathbb{P}_\theta$ from its surrogate posterior and measure $\mathbb{Q}_\xi$. There is in othert words, no need for variational inference since the latent states are directly observed. This impacts our loss formulation as discussed in \ref{sec:dynROM}. Crucially, on a more conceptual note, we use the SDE to model the inherent stochasticity of the chaotic system itself, rather than merely to quantify predictive uncertainty as in \citep{Tornado, DDStoROM, SDE-net}. Aside from a few exceptions, such as \citep{StochasticMZ}, few reduced-order models, to our knowledge, explicitly treat macroscopic systems, and chaotic systems in particular, as stochastic processes, even-though these systems are known to exhibit a form of irreducible amount of stochasticity, passed their Lyapunov time scale. This work draws inspiration from \citep{Neural-SDE}, which follows a similar conceptual path, modelling complex dynamical systems that feature bifurcations, multimodality, and sharp transitions. However, it does not address chaotic dynamics or non-Markovian effects, both of which we aim to incorporate by extending this methodology.

\section{Methodology}

\label{sec:dynROM}

We consider the time-dependant state $\varphi_t$ of a high-dimensional dynamical system characterized by \textit{'ch'} channels or fields and a $n_x \times n_y$ spatial resolution.
\[
\varphi_t \in \mathbb{R}^{\mathrm{ch} \times n_x \times n_y},
\]
The lower-dimensional manifold is identified using an autoencoder composed of an encoder $\mathcal{E}$ and a decoder $\mathcal{D}$ such that
\[
\mathcal{D} \circ \mathcal{E} : \varphi \mapsto \varphi.
\]
The identified low-dimensional latent space $Z \subset \mathbb{R}^d$, has the dimension $d \ll \mathrm{ch} \times n_x \times n_y$, defined as
\[
\mathcal{E} : \mathbb{R}^{\mathrm{ch} \times n_x \times n_y} \to \mathbb{R}^d, 
\quad \varphi \mapsto z.
\]
The decoder $\mathcal{D}$ then maps the latent variables back to the original high-dimensional space,
\[
\mathcal{D} : \mathbb{R}^d \to \mathbb{R}^{\mathrm{ch} \times n_x \times n_y},
\quad z \mapsto \hat{\varphi},
\]
where, $\hat{\varphi}$ denotes the reconstruction of the input state. The latent variable $(z_t)_{t \geq 0}$ are evolved by a stochastic process in $\mathbb{R}^d$. At each time $t$, $z_t$ is a random variable whose distribution characterizes the uncertainty in the latent representation. In this sense, the latent dynamics can be viewed as the flow of a probability distribution in the latent space.

This probability is then modeled using an Itô-type SDE (Eq. \ref{eq:SDE}) to capture two distinct sources of stochasticity: the system's inherent chaotic dynamics and multi-admissible futures, as well as the stochasticity or 'noise' stemming from lost information in the projection's orthogonal subspace \citep{Mori, Zwanzig}, following
\begin{equation}
\label{eq:SDE}
\mathrm{d}z_t = f(z_t)\,\mathrm{d}t +\gamma(z_t)\,\mathrm{d}W_t,
\end{equation}
where $f : \mathbb{R}^d \to \mathbb{R}^d$ is the drift function encoding the deterministic component of the dynamics, $\gamma : \mathbb{R}^d \to \mathbb{R}^{d \times d}$ is the diffusion function encoding stochastic effects, and $(W_t)_{t \geq 0}$ is a standard $d$-dimensional Wiener process. This formulation induces a family of transition kernels $(p_t)_{t \geq 0}$ on the latent space, where $p_t(z \mid z_0)$ denotes the conditional distribution of $z_t$ given the initial state $z_0$. Under standard regularity assumptions namely continuity and Lipschitz-continuity on $f$ and $\gamma$, these transition kernels are Gaussian in the infinitesimal sense. Its further assumed that the coefficients $f$ and $\gamma$ are time-homogeneous, i.e., they do not depend explicitly on time. Any implicit time dependence arises solely through the state $z_t$. Finally, the stochastic components of the latent variables are assumed to remain conditionally independent throughout the temporal evolution, which amounts to restricting the diffusion matrix to be diagonal, so that
\[
\gamma(z) = \mathrm{diag}(g(z)), \quad \text{with } g : \mathbb{R}^d \to \mathbb{R}^d.
\]
Equivalently, the SDE can be written component-wise as
\begin{equation}
\label{eq:element_wise_SDE}
\mathrm{d}z_t^{(i)} = f_i(z_t)\,\mathrm{d}t + g_i(z_t)\,\mathrm{d}W_t^{(i)}, 
\quad i = 1, \dots, d,
\end{equation}
where, the $(W_t^{(i)})_{i=1}^d$ are independent standard Wiener processes.

In a discrete-time setting, the above continuous SDE can be approximated using the Euler-Maruyama scheme. For a sufficiently small time step $\Delta t > 0$, such that the infinitesimal Gaussian approximation of the transition kernel remains valid, we obtain
\begin{equation}
\label{eq:Euler-Mayurama}
z_{t+\Delta t} = z_t + f(z_t)\,\Delta t + g(z_t) \odot \Delta W_t,
\end{equation}
where,
\[
\Delta W_t \sim \mathcal{N}(0, \Delta t\, I_d) .
\]
For simplicity, a constant time step $\Delta t = 1$ is assumed, interpreted as a unit increment required to be sufficiently small for the infinitesimal Gaussianity approximation to remain valid. This yields the discrete-time dynamical system
\begin{equation}
\label{eq:simplified_SDE}
z_{t+1} = z_t + f(z_t) + g(z_t) \odot \xi_t,
\end{equation}
where,
\[
\xi_t \sim \mathcal{N}(0, I_d), \quad \text{i.i.d.}
\]

In the data-driven setting adopted here, the drift and diffusion coefficients are parameterized by neural networks $f_\theta$ and $g_\theta$. The learned latent dynamics are then given by
\begin{equation}
\label{eq:data-driven-SDE}
z_{t+1} = z_t + f_\theta(z_t) + g_\theta(z_t) \odot \xi_t,
\end{equation}
with $\xi_t \sim \mathcal{N}(0, I_d)$. There is, however, a fundamental limitation to this formulation in its current form: the proposed Gaussian transition kernel is Markovian, which is likely to prevent the model from learning any meaningful drift component. Indeed, such dynamical systems — and their latent representations in particular, are often non-Markovian \cite{MZAE}. It is therefore essential to incorporate a memory term, as supported by the Mori–Zwanzig formalism, in order to recover, at least partially, the dynamics lost in the orthogonal complement of the projection subspace. From a probabilistic perspective, this implies that the transition kernel is path-dependent:
\[
p(z_{t+1} \mid z_t, z_{t-1}, \dots).
\]
To account for such memory effects, an augmented latent space is used instead, 
\[
Y \subset \mathbb{R}^{d \times \tau},
\]
where $\tau \in \mathbb{N}$ denotes the length of a finite look-back window. The augmented state $y_t \in Y$ is defined as
\[
y_t =
\begin{bmatrix}
z_t \\
z_{t-1} \\
\vdots \\
z_{t-\tau+1}
\end{bmatrix}.
\]
By construction, the process $(y_t)_{t\geq 0}$ is Markovian, even if $(z_t)_{t\geq 0}$ is not. As a result, the augmented dynamics can be modeled using a Gaussian Markov transition,
\begin{equation}
\label{eq:Augmented_SDE}
y_{t+1} = y_t + \mathcal{F}_\theta(y_t) + \mathcal{G}_\theta(y_t) \odot \xi_t,
\end{equation}
\begin{center}
with,   
$
y_{t+1} = 
\begin{bmatrix}
z_{t+1} \\ z_t \\ \vdots \\ z_{t-\tau+2}
\end{bmatrix}$,
$\mathcal{F}_\theta = \begin{bmatrix}
f_\theta(y_t) \\
z_t - z_{t-1} \\
\vdots \\
z_{t-\tau+2} - z_{t-\tau+1}
\end{bmatrix} \in \mathbb{R}^{d\times \tau}$,  $\mathcal{G}_\theta = \begin{bmatrix}
g_\theta(y_t) \\
0 \\
\vdots \\
0
\end{bmatrix} \in \mathbb{R}^{d\times \tau}$ and 
$\xi_t \in \mathbb{R}^{d\times \tau}$,
\end{center}
which propagates the memory window forward in time. 

This formulation enables the likelihood of observed transitions to be evaluated under the proposed model, which is in turn used to determine the parameters of $f_\theta$ and $g_\theta$. Given the Gaussian Markov transition kernel, the conditional distribution of the augmented state $y_{t+1}$ given $y_t$ is
\[
p(y_{t+1} \mid y_t)
=
\mathcal{N}\!\left(
y_t + \mathcal{F}_\theta(y_t),\;
\mathrm{diag}\!\big(\mathcal{G}_\theta^2(y_t)\big)
\right),
\]
where the covariance matrix is supported only on the first $d$ components and is given by
\[
\mathrm{diag}\!\big(\mathcal{G}_\theta^2(y_t)\big)
=
\begin{bmatrix}
\mathrm{diag}\!\big(g_\theta^2(y_t)\big) & 0 \\
0 & 0
\end{bmatrix}
\in \mathbb{R}^{d\tau \times d\tau}.
\]
This defines a degenerate Gaussian Markov kernel on the augmented space, with stochasticity restricted to the leading $d$ coordinates. As a consequence, the transition can be equivalently characterized by its marginal on the first $d$ components:
\[
p(y_{t+1} \mid y_t) = p(z_{t+1} \mid y_t),
\]
with
\[
p(z_{t+1} \mid y_t)
=
\mathcal{N}\!\left(
z_t + f_\theta(y_t),\;
\mathrm{diag}\!\big(g_\theta^2(y_t)\big)
\right).
\]
This equivalence follows from the fact that only the first $d$ components of $y_{t+1}$ are stochastic (namely $z_{t+1}$), while the remaining $(\tau - 1)d$ components are deterministic shifts of the history window.

\subsection{Latent learning objective}
The parameters of $f_\theta$ and $g_\theta$ are estimated by minimizing the expected negative log-likelihood (NLL) over a dataset $\mathcal{X}$ of transitions:
\begin{equation}
\label{eq:objective}
f_\theta,\, g_\theta 
= \arg\min_\theta \; \mathbb{E}_{(y_t,\, z_{t+1})\,\sim\,\mathcal{X}}\!\left[\mathcal{L}_{z_{t+1} \mid y_t}\right].
\end{equation}
Dropping constant terms, the NLL of a single transition is given by
\begin{equation}
\label{eq:NLL}
\mathcal{L}_{z_{t+1} \mid y_t} 
=
-\log p(z_{t+1} \mid y_t)
=
\frac{1}{2}\sum_{i=1}^{d}
\left[
\frac{\bigl(z_{t+1}^{(i)} - z_t^{(i)} - f_i(y_t)\bigr)^2}{g_i^2(y_t)}
+ \log g_i^2(y_t)
\right].
\end{equation}
Using the NLL as a training objective offers two complementary advantages. First, it explicitly accounts for the uncertainty arising from the latent system's irreducible uncertainty. Second, it enables the time-integrator to compensate for inaccurate drift estimates through increased diffusion, thereby capturing uncertainties inherent to the temporal integration itself.

\begin{figure}[H]
    \centering
    \includegraphics[width=1\linewidth]{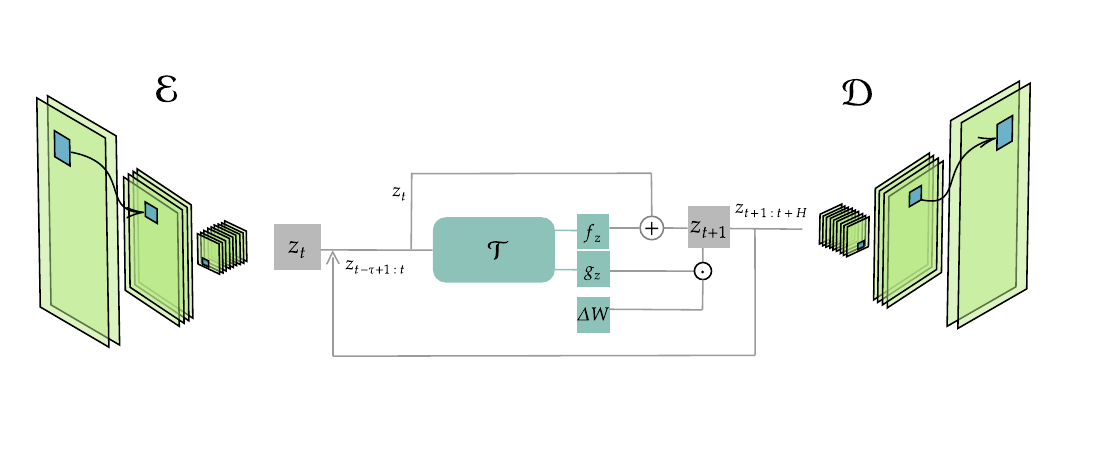}
    \caption{Data pipeline over a prediction rollout of H steps using $\tau$ steps of lookback window.}
    \label{fig:data-pipeline}
\end{figure}

Both $f_\theta$ and $g_\theta$ are jointly parametrized by a single transformer $\mathcal{T}$,  leveraging the state-of-the-art performance of attention mechanisms for sequence forecasting \citep{AttentionAllYouNeed}:
\[
\mathcal{T} : y_t \;\mapsto\; f_\theta(y_t),\; g_\theta(y_t).
\]
The model is trained over a fixed rollout horizon of H steps, during which predicted states are fed back as input context, forcing the model to learn to propagate dynamics auto-regressively rather than relying solely on ground-truth observations. The pipeline is illustrated in Figure \ref{fig:data-pipeline}.

Note that during auto-regressive rollout, predictions are conditioned on the predicted augmented state \(\hat{y}_t\) rather than the true state \(y_t\), yielding a learned surrogate transition kernel
\[
p_\theta(z_{t+1}\mid \hat{y}_t).
\]
Let \(\mathbb{Q}_\theta\) denote the model-induced rollout distribution. The training objective after many rollout steps is the corresponding on-policy negative log-likelihood:
\begin{equation}
\label{eq:auto-reg_Nll}
\mathcal{L}_{\text{NLL}}(\theta)
=
\mathbb{E}_{(z_{t+1},\,\hat{y}_t)\sim \mathbb{Q}_\theta}
\big[-\log p_\theta(z_{t+1}\mid \hat{y}_t)\big].
\end{equation}
with : 
\begin{equation}
\label{eq:auto-reg_Nll-details}
-\log p(z_{t+1} \mid \hat{y_t})
=
\frac{1}{2}\sum_{i=1}^{d}
\left[
\frac{\bigl(z_{t+1}^{(i)} - \hat{z}_t^{(i)} - f_i(\hat{y_t})\bigr)^2}{g_i^2(\hat{y_t})}
+ \log g_i^2(\hat{y_t})
\right].
\end{equation}

The transition sampled from the true rollout distribution \(\mathbb{P}\), conditioned on a model-generated path \(\hat{y}_t\), given by \begin{equation}
\label{eq:true_kernel_nll}
\mathbb{E}_{(z_{t+1},\,\hat{y}_t)\sim \mathbb{P}, \;\;\;
\hat{y_t}\sim \mathbb{Q}_\theta}
\big[-\log p_\theta(z_{t+1}\mid \hat{y}_t)\big].
\end{equation}
is intractable, unless one somehow integrates the DNS solver in the training routine to query it at every model rollout step. We therefore optimize the surrogate objective \(\mathcal{L}(\theta)\) from Equation \ref{eq:auto-reg_Nll} as a tractable approximation. One could also use teacher forcing, i.e. conditioning on true data, to optimize the objective:
\begin{equation}
\label{eq:teacher_forcing}
\mathbb{E}_{(z_{t+1},\,y_t)\sim \mathbb{P}}
\big[-\log p_\theta(z_{t+1}\mid y_t)\big].
\end{equation}
However, this latter objective breaks the auto-regressive conditioning, and the model would lose its generative properties. The training routine is given in its pseudo-code format in alg. \ref{alg:latent_sde_training}.

\subsection{Training}
The autoencoder $(\mathcal{E,D})$ and the latent integrator $\mathcal{T}$ are trained jointly so that the learnt latent representation also benefits the dynamical integration. The resulting objective function can be written as  
\begin{equation}
\label{eq:loss_dynROM}
\mathcal{L}_{total}
=
\mathbb{E}_{z \sim q(z \mid \varphi)}
\left[
\left\| \varphi - \mathcal{D}(z) \right\|^2
\right]
+ 
\alpha \, \mathbb{E}_{(z_{t+1},\,\hat{y}_t)\sim \mathbb{Q}_\theta}
\left[
-\log p_\theta(z_{t+1}\mid \hat{y}_t)
\right],
\end{equation}
with $q$ defining the encoder distribution over all input states $\varphi$. In the context of classical, deterministic Autoencoders, the conditional distribution $q(z \mid \varphi)$ is a Dirac function, $\delta(z)$. The first term in the above objective function corresponds to the reconstruction loss, enforcing that the decoded signal remains close to the input. The second term is the latent integration objective calculated with equation \ref{eq:auto-reg_Nll}. Architecture specifications, detailed loss and hyperparametrs are provided in the Appendix \ref{subsec:app-arch_loss}. 

Our latent dynamics loss (Equation~\ref{eq:NLL}) can be recovered as a special case of the general SDE matching objective (Equation~\ref{eq:loss_SDEmatching}) under the discrete-time, deterministic-autoencoder, diagonal-diffusion limit discussed above. However, one structural difference remains: whereas $\mathcal{L}_{\text{diff}}$ in Equation~\ref{eq:loss_SDEmatching} reduces to a pure drift-residual term, our objective retains an explicit $\log g_i^2(y_t)$ normalization term. That is because the Girsanov-based derivation of $\mathcal{L}_{\text{diff}}$ relies on the posterior and prior path measures sharing an identical diffusion coefficient $g_\theta$, which causes the Gaussian normalizing constants of both measures to cancel exactly in the resulting KL divergence (the two measures $\mathbb{P}_\theta$ and $\mathbb{Q}_\xi$ mutually regularize each other). In our setting, by contrast, there is no separate reference process to cancel against: $p_\theta(z_{t+1} \mid y_t)$ is fit directly to observed transitions via maximum likelihood, so its regularization is retained.

\begin{algorithm}[H]
\caption{Training of latent dynamics model}
\label{alg:latent_sde_training}
\begin{algorithmic}[1]
\Require Dataset $\{\varphi_t\}_{t=1}^T$, window size $\tau$, horizon $H$, regularization weight $\alpha$
\Ensure Encoder $\mathcal{E}$, decoder $\mathcal{D}$, dynamics model $\mathcal{T}$

\For{each epoch}
  \For{each minibatch centered at time $t$}
  
    \State \textbf{Extract sequences:}
    \State $\{\varphi_s\}_{s=t-\tau+1}^{t}$
    \State $\{\varphi_s\}_{s=t+1}^{t+H}$
    
    \State \textbf{Encode past:}
    \State $\{z_s\}_{s=t-\tau+1}^{t} \gets \mathcal{E}(\{\varphi_s\}_{s=t-\tau+1}^{t})$
    
    \State \textbf{Encode future (target):}
    \State $\{z_s\}_{s=t+1}^{t+H} \gets \mathcal{E}(\{\varphi_s\}_{s=t+1}^{t+H})$

    \State Initialize context $z_{\mathrm{ctx}} \gets \{z_s\}_{s=t-\tau+1}^{t}$
    \State Initialize predictions $\{\hat{z}_{t+n}\}_{n=1}^{H} \gets \emptyset$
    
    \For{$n = 1, \dots, H$}
        \State $(f_{t+n}, g_{t+n}) \gets \mathcal{T}(z_{\mathrm{ctx}})$
        \State Sample $\xi_n \sim \mathcal{N}(0, I)$
        
        \State $z_{\mathrm{last}} \gets \text{last}(z_{\mathrm{ctx}})$
        
        \State $\hat{z}_{t+n} \gets 
        z_{\mathrm{last}} + f_{t+n} + g_{t+n} \odot \xi_n$
        
        \State $\{\hat{z}_{t+n}\}_{n=1}^{H} \gets \text{append}(\{\hat{z}_{t+n}\}_{n=1}^{H}, \hat{z}_{t+n})$
        
        \State $z_{\mathrm{ctx}} \gets 
        \text{concat}(z_{\mathrm{ctx}}[2:\ ], \hat{z}_{t+n})$
    \EndFor
    
    \State \textbf{Reconstruction loss:}
    \[
    \mathcal{L}_{\mathrm{recon}} \gets
    \frac{1}{\tau}
    \sum_{s=t-\tau+1}^{t}
    \mathbb{E}_{q(z_s \mid \varphi_s)}
    \big[
    \|\varphi_s - \mathcal{D}(z_s)\|^2
    \big]
    \]
    
    \State \textbf{Latent dynamics loss:}
    \[
    \mathcal{L}_{\mathrm{latent}} \gets
    \mathcal{L}_{\mathrm{NLL}}\Big(
    \{z_{t+n}\}_{n=1}^{H},
    \{f_{t+n}, g_{t+n}\}_{n=1}^{H}
    \Big)
    \]
    
    \State \textbf{Total loss:}
    \[
    \mathcal{L} \gets \mathcal{L}_{\mathrm{recon}} + \alpha\, \mathcal{L}_{\mathrm{latent}}
    \]
    
    \State Update parameters of $\mathcal{E}, \mathcal{D}, \mathcal{T}$
    
  \EndFor
\EndFor

\State \Return $\mathcal{E}, \mathcal{D}, \mathcal{T}$
\end{algorithmic}
\end{algorithm}

\section{Flow configurations}
\label{sec:experiments}

The proposed approach is validated on two standard PDE benchmarks known to exhibit chaotic dynamics: the two-dimensional Navier–Stokes equations configured as a Kolmogorov flow, and the one-dimensional Kuramoto–Sivashinsky equation. These flow configurations are briefly described below.

\subsection{Kolmogorov flow}
\label{subsec:exp_kolmo}
One of the benchmarks considered in this work is the \textit{Kolmogorov flow}, a two-dimensional, incompressible, and unsteady solution of the Navier-Stokes equations defined on a doubly periodic domain,
\[
(x,y) \in [0,2\pi] \times [0,2\pi].
\]
The flow is sustained by a spatially periodic body force acting in the $x$-direction. The governing equations are given by
\begin{equation}
\label{NS}
\begin{aligned}
\frac{\partial \mathbf{u}}{\partial t}
    &= -\nabla p - (\mathbf{u} \cdot \nabla)\mathbf{u} + \frac{1}{Re}\nabla^2 \mathbf{u} + \mathbf{f}, \\
\nabla \cdot \mathbf{u}
    &= 0, \\
\mathbf{f}(x,y)
    &= \sin(n_f y)\,\mathbf{e}_x,
\end{aligned}
\end{equation}
where $\mathbf{u} = (u,v)$ denotes the velocity field, $p$ the pressure, $Re$ the Reynolds number, and $\mathbf{f}$ the external forcing. The forcing term injects energy at a prescribed spatial scale through a sinusoidal excitation of wavenumber $n_f$.

Kolmogorov flow provides a canonical setting for studying  turbulence~\citep{Kolmogorov}. Depending on the choice of $Re$ and $n_f$, the system exhibits a wide range of dynamical regimes, from laminar states to fully developed turbulence~\citep{KolmogorovRe_Nf}. At sufficiently high Reynolds numbers, the flow develops an energy cascade in which energy injected at large scales is transferred to progressively smaller scales and ultimately dissipated by viscosity. In this study, we consider $Re = 90$ and forcing wavenumber $n_f = 4$. The characteristic length $L$ entering the Reynolds number, $Re = \frac{\rho U_\infty L}{\mu}$, is defined by the sinusoidal forcing amplitude. At such a Reynolds number, the flow exhibit chaotic and multi-scale behaviour.

\begin{figure}[H]
    \centering
    \begin{minipage}{0.48\textwidth}
        \begin{subfigure}{\textwidth}
            \includegraphics[width=1\linewidth]{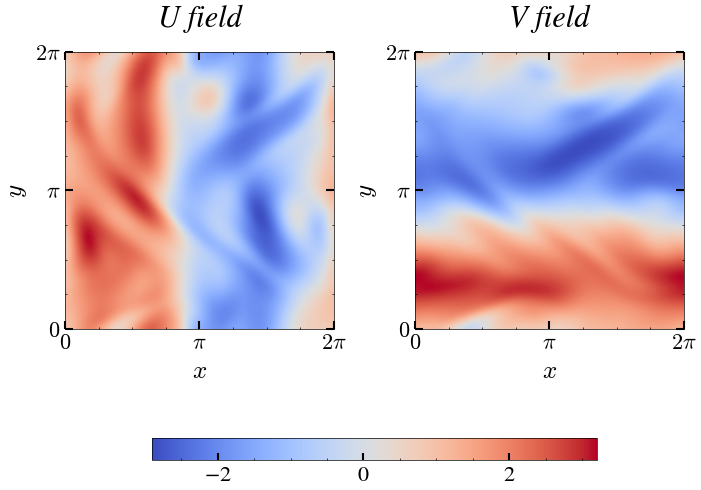}
            \caption{Velocity fields $U, V$}
        \end{subfigure}
    \end{minipage}
    \hfill
    \begin{minipage}{0.48\textwidth}
        \centering
        \begin{subfigure}{\textwidth}
            \includegraphics[width=0.9\linewidth]{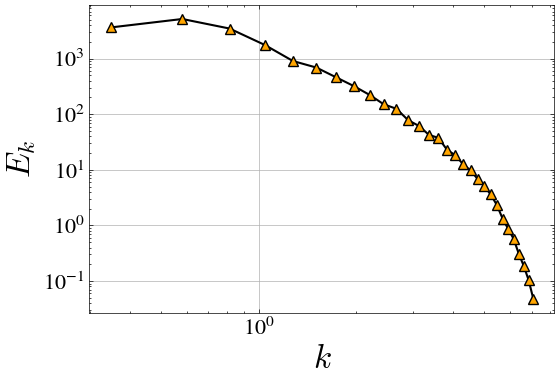}
            \caption{Kolmogorov energy cascade}
        \end{subfigure}
    \end{minipage}
    \caption{Snapshots of the Kolmogorov velocity fields $U$ and $V$ (left) and the corresponding energy cascade (right).}
    \label{fig:Kolmogorovflow}
\end{figure}

The flow is simulated on a $64 \times 64$ uniform grid using the publicly available pseudo-spectral solver \textit{kolSol}~\citep{Canuto}. The solution is represented using $32$ Fourier modes in each spatial direction. Time integration is performed in Fourier space with a time step $\delta t = 0.01$. Snapshots of the velocity field are recorded every $\delta t = 0.12$. The resulting dataset consists of velocity fields $\mathbf{u}(x,y,t),\space \mathbf{v}(x,y,t)$ stored as tensors in $\mathbb{R}^{n_x \times n_y \times T}$ with $n_x = n_y = 64$ and $T = 1000$ time steps per trajectory. An ensemble of $10$ independent trajectories is generated.
Figure~\ref{fig:Kolmogorovflow} shows the energy cascade alongside one snapshot of the Kolmogorov velocity fields, \(U\) and \(V\).

For the learning task, the dataset is split along the temporal dimension, with $80\%$ of the snapshots used for training and the remaining $20\%$ for testing. This results in a training set 
\[
\mathcal{X} \in \mathbb{R}^{10 \times T_{\text{train}} \times 2 \times n_x \times n_y}, \quad T_{\text{train}} = 800,
\]
and a test set
\[
\mathcal{X}_{\text{test}} \in \mathbb{R}^{10 \times T_{\text{test}} \times 2 \times n_x \times n_y}, \quad T_{\text{test}} = 200,
\]
with $(n_x,n_y) = (64,64)$.

\subsection{Kuramoto-Sivashinsky (1D-KS)}
\label{subsec:exp_ks}

We also validate our methodology with a 1-D canonical system. The Kuramoto-Sivashinsky (KS) equation is a fourth-order nonlinear PDE widely utilized as a benchmark in turbulence modeling and the study of spatiotemporal chaos. It is defined as:
\begin{equation}
\frac{\partial u}{\partial t} + u \frac{\partial u}{\partial x} + \frac{\partial^2 u}{\partial x^2} + \nu \frac{\partial^4 u}{\partial x^4} = 0,
\end{equation}
where, $u(x,t)$ denotes the scalar field of interest, $x$ is the spatial coordinate, $t$ represents time, and $\nu > 0$ is the viscosity parameter. The KS equation captures the essential features of chaos, namely, sensitive dependence on initial conditions and complex spatio-temporal patterns, while remaining computationally tractable. This makes it an ideal test-bed for evaluating data-driven PDE emulators in chaotic regimes. 

\begin{figure}[H]
    \centering
    \includegraphics[width=1\linewidth]{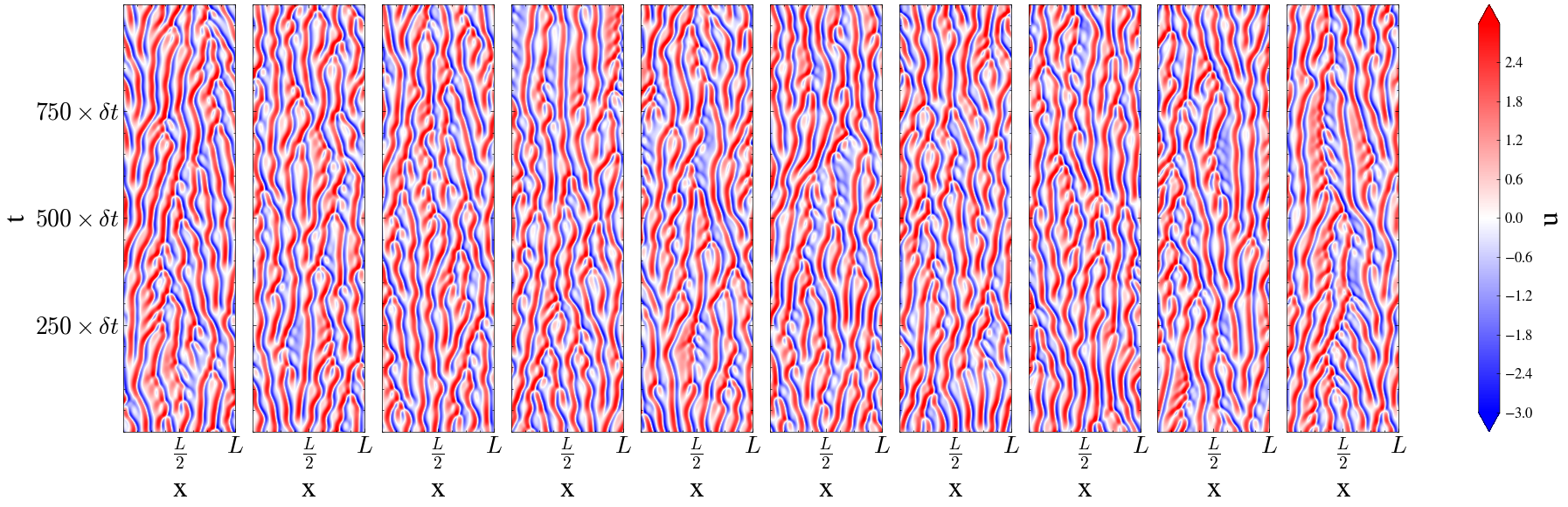}
    \caption{Numerical integration of 10 independent 1D-KS trajectories over the time horizon $T$, each starting from a unique initial condition.}
    \label{fig:KS}
\end{figure}

The 1D Cauchy problem for the KS equation is well-posed, and for a given initial condition $u(x,0)$, there exists a unique solution $u(x,t)$ for $t \in [0, \infty]$ that depends continuously on the initial data. However, long-term forecasting remains highly challenging due to the system's chaotic nature, where infinitesimal initial errors amplify exponentially over time, eventually degrading emulator performance. In this work, we fix $\nu = 0.8$ to ensure the system operates within this characteristically chaotic regime. Numerical simulations are performed over a spatial domain of fixed size $L = 512\delta x$ subject to periodic boundary conditions. The system is integrated over a time horizon $T = 1000\delta t$ with a dimensionless time increment of $\delta t = 0.25$. To expose the emulator to a diverse range of dynamical scenarios, thereby leveraging their stochastic properties to explore multiple admissible futures under identical past-conditioning, we generate an ensemble of 10 distinct trajectories differing solely in their initial conditions. This approach yields a rich variety of spatio-temporal transitions, illustrated as $(x,t)$ plots in Figure \ref{fig:KS}. The reasoning that motivates the number of trajectory used is detailed with an ablation study in \ref{Appendix:ablation}.

For the learning task, the dataset is split along the temporal dimension, with $80\%$ of the snapshots used for training and the remaining $20\%$ for testing. This results in a training set 
\[
\mathcal{X} \in \mathbb{R}^{10 \times T_{\text{train}} \times n_x}, \quad T_{\text{train}} = 800,
\]
and a test set
\[
\mathcal{X}_{\text{test}} \in \mathbb{R}^{10 \times T_{\text{test}} \times n_x}, \quad T_{\text{test}} = 200,
\]. 

Throughout this work, predicted quantities are indicated using a hat notation \((\,\hat{\cdot}\,)\).
\section{Results}
\label{sec:results}
\subsection{2D - Kolmogorov flow}
\label{subsec:results_kolmo}

\begin{figure}[H]
    \centering
    \includegraphics[width=0.6\linewidth]{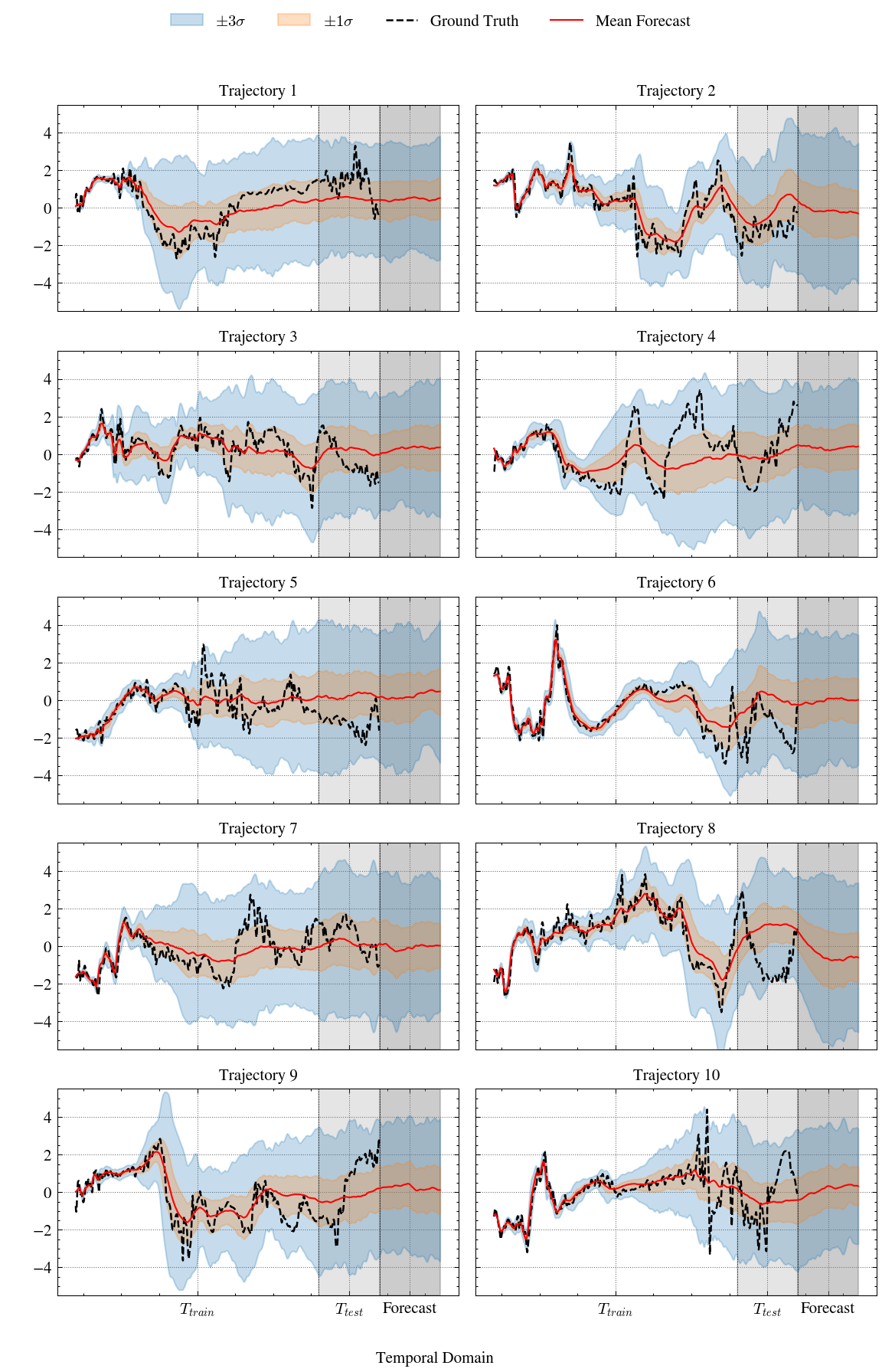}
    \caption{Reconstruction of the velocity component $U_{x_i,y_i}(t)$ at the spatial location $(x_i,y_i)=(\pi,\pi)$ over an autoregressive rollout horizon of $T = 1200\,\delta t$. For each initial condition, an ensemble of $60$ stochastic trajectories is generated by the model. The solid line denotes the ensemble mean, while the shaded regions represent the intervals corresponding to one and three standard deviations around the mean, illustrating the uncertainty over time.}
    \label{fig:10_traj_time_Kolmo}
\end{figure}

To assess the model’s ability to reproduce and extrapolate the flow dynamics, we initialize the model from the same $10$ training initial conditions and roll out its predictions over an extended horizon of $1200$ snapshots. This evaluation window therefore includes the training horizon, the testing horizon, and an additional forecasting horizon of $200$ snapshots beyond the available data. This procedure is repeated over a defined ensemble cardinality to evaluate the stochastic properties of the model. The results of that evaluation are compiled in Figure \ref{fig:10_traj_time_Kolmo} at a given spatial location $(x, y) \in[0, 2\pi]^2$.
The results shown in Figure~\ref{fig:10_traj_time_Kolmo} indicate that the stochastic trajectories accurately follow the ground-truth trajectory from a given initial condition over a finite and variable time horizon, before eventually diverging, often before the test window. While the MSE with the mean trajectory remains moderate on the training set, it increases substantially on the test set, reaching the same order of magnitude as the signal itself as supported by table \ref{tab:MSE_PICP_kolmo}. This is accompanied by an increase in predictive standard deviation, indicating that each stochastic realization follows its own trajectory and becomes progressively less correlated with the ground truth over time. Such behaviour is characteristic of chaotic systems, where small perturbations accumulate and induce trajectory separation. 

\begin{table}[H]
\centering
\caption{Evaluation Metrics across Trajectories}
\label{tab:MSE_PICP_kolmo}
\begin{tabular}{lccc}
\toprule
 & MSE & PICP $1\sigma$ & PICP $3\sigma$ \\
\midrule
Train Set & 0.53 & 61\% & 93\% \\
Test Set  & 1.70 & 59\% & 92\% \\ 
\bottomrule
\end{tabular}
\end{table}

Although the distance between the ensemble mean and the ground truth increases over the prediction horizon, the prediction interval coverage probability (PICP) remains relatively stable as supported by table \ref{tab:MSE_PICP_kolmo} suggesting that the model's uncertainty weights out the mean discrepancy. The PICP is defined by the expression given in \ref{eq:PICP} 
\begin{equation}
\label{eq:PICP}
\mathrm{PICP}_{k\sigma} = \frac{1}{N}\sum_{i=1}^{N} \mathbb{I}\left( y_i \in \left[\mu_i - k\sigma_i,\; \mu_i + k\sigma_i\right]\right),
\end{equation}
where, \(N\) is the number of prediction points, \(y_i\) is the ground-truth value, \(\mu_i\) the ensemble mean prediction, \(\sigma_i\) the predictive standard deviation, and \(\mathbb{I}(\cdot)\) the indicator function.

An important observation is that the distance between the ensemble mean and the ground truth does not blow up and often remains within the same range even after individual realizations have significantly decorrelated. This should, however, be interpreted cautiously: the ensemble mean does not correspond to a physically realizable trajectory and does not necessarily satisfy the learned dynamics. In high-variance regimes, the mean may lie in regions of state space that are unlikely to be visited by any individual trajectory. We therefore use it primarily as an indicator that the probability mass remains centered around the reference trajectory. This is why individual realizations can provide more dynamical insight than the ensemble mean.

\begin{figure}[H]
    \centering
    \includegraphics[width=1\linewidth]{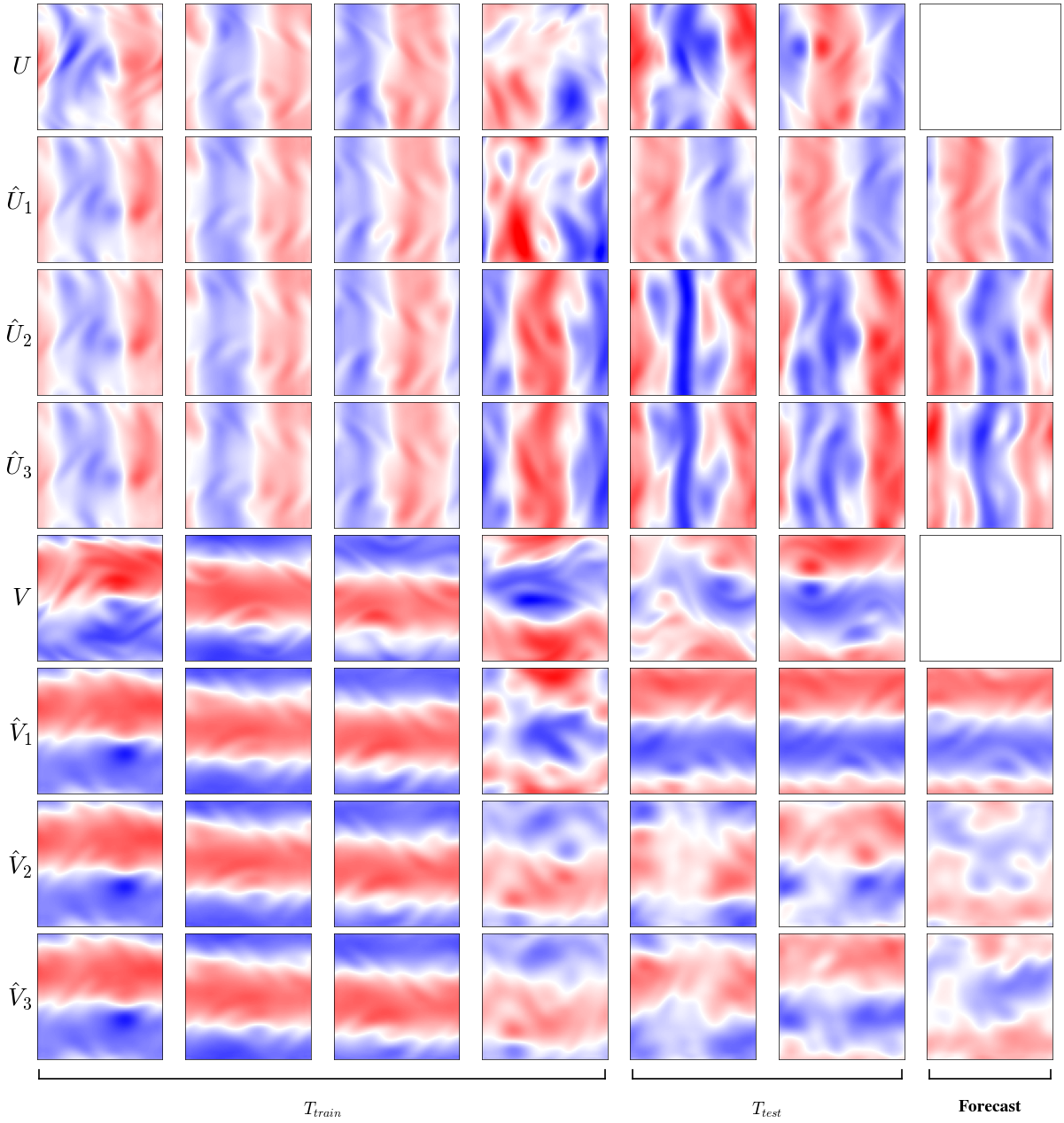}
    \caption{Comparison between a reference trajectory ($U_{t_i}(x,y)$ top, $V_{t_i}(x,y)$ bottom) and three stochastic realizations $(\hat{U_j}, \hat{V_j})_{j=1,2,3}$ initialized from the same initial condition. The rollout goes passed the training window, with shown snapshots $t_i\in \{1, 250, 500, 750\} \times \delta t$, and the test window with shown snapshots: $t_i \in \{850, 950\} \times \delta t$. Finally a forecast window is also represented at the snapshot : $t_i = 1100 \times \delta t$. Snapshots were selected arbitrarily.}
    \label{fig:10_traj_space_kolmo}
\end{figure}

Figure~\ref{fig:10_traj_space_kolmo} compares the evolution of three trajectories generated by the model from the same initial condition, alongside the corresponding reference trajectory. The figure highlights that, prior of the testing window, trajectories begin to diverge and explore distinct regions of state space. While these trajectories separate from the reference and from each other, their evolution remains dynamically plausible, illustrating the model’s ability to capture multiple admissible futures under stochastic dynamics.

\begin{figure}[H]
    \centering
    \includegraphics[width=0.8\linewidth]{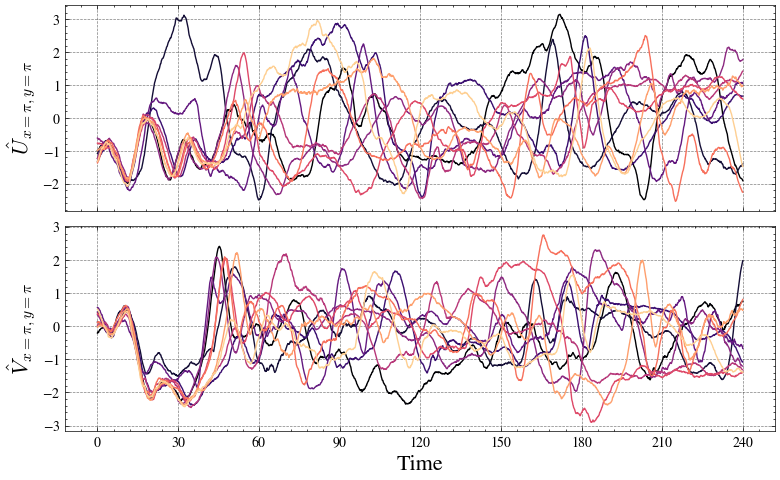}
    \caption{Temporal evolution of the velocity components (\(U_{x_i, y_i}(t)\) top, \(V_{x_i, y_i}(t)\) bottom) at the spatial location \((x_i,y_i)=(\pi,\pi)\) for 10 generated trajectories over a rollout of 2000 snapshots. For visualization purposes only, the trajectories are smoothed using a moving-average convolution with a sliding window spanning $40 \delta t$.}
    \label{fig:new_10_traj_time_kolmo}
\end{figure}

To assess the learned transition kernel beyond trajectory reconstruction, we evaluate its generative capability and stability under long-term rollouts from an unseen initial condition. The goal is to verify that trajectories sampled from the model remain dynamically consistent with the underlying system and stable as the rollout progresses, rather than simply reproducing trajectories observed during training by "memorizing them". To this end, we generate an ensemble of 10 trajectories initialized from a synthetic initial condition outside the training set. Each trajectory is obtained by recursively sampling from the learned transition kernel at every rollout step over 2000 snapshots thus largely exceeding the train and test window encompassing a thousand snapshots. Emulators of chaotic PDEs often accumulate errors auto-regressively, leading to exponential error growth and unphysical realizations after only a few rollout steps in the absence of explicit stabilization \citep{Thermalizer, ESN}. Crucially, the framework proposed in this work must maintains stability well beyond the training horizon.  
Figure~\ref{fig:new_10_traj_time_kolmo} shows the temporal evolution of $\hat{U}_{x_i, y_i}(t)$ at a given spatial location $(x_i, y_i)$. This set of 10 generated trajectories is now the subject of evaluation until the end of the this subsection. Importantly, no ground-truth reference trajectory is considered in this experiment. The objective is not to evaluate the model’s ability to reproduce a known path, but rather to assess whether the generated trajectories remain dynamically plausible and stable under long-term autonomous rollout.

\begin{figure}[H]
    \centering
    \includegraphics[width=0.9\linewidth]{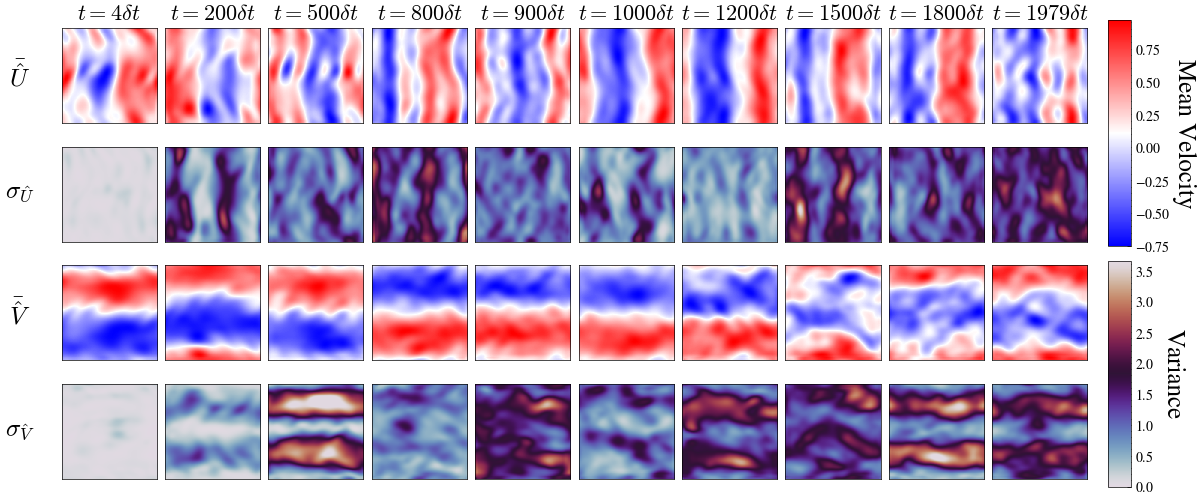}
    \caption{Mean and standard deviation of $\hat{U}$ and $\hat{V}$ throughout the prediction horizon.}
    \label{fig:new_10_traj_space_kolmo}
\end{figure}

Figure~\ref{fig:new_10_traj_space_kolmo} shows the first and second moments of the generated ensemble over the prediction horizon. The variance remains bounded and does not simply accumulate over time. Rather, it varies both temporally and spatially, with uncertainty appearing in localized regions of the state space causing trajectories to diverge and explore different regions of the attractor. 

\begin{figure}[H]
    \centering
    
    \begin{subfigure}[b]{0.365\linewidth}
        \centering
        \includegraphics[width=\linewidth]{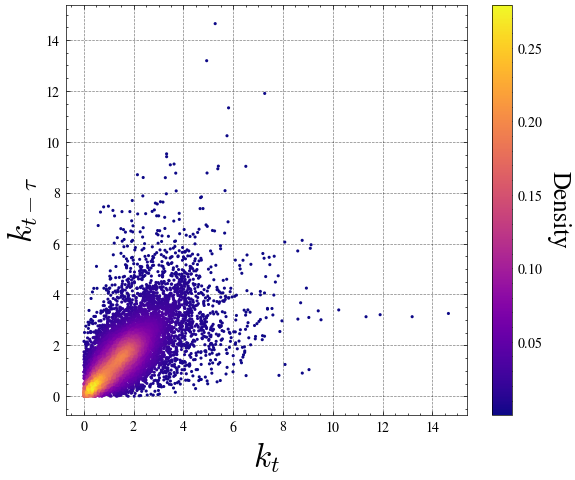}
        \caption{Reference manifold obtained from the kinetic energy delay embedding.}
        \label{fig:energy_delay_embedding}
    \end{subfigure}
    \hfill
    \begin{subfigure}[b]{0.60\linewidth}
        \centering
        \includegraphics[width=\linewidth]{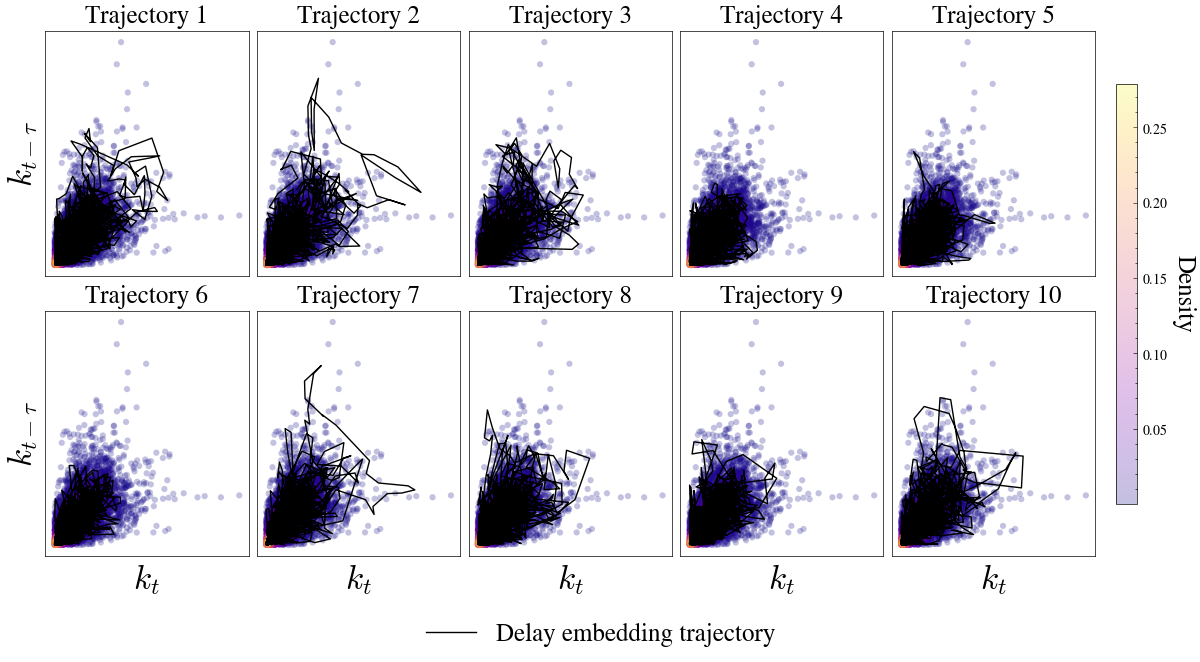}
        \caption{Delay embeddings of the generated stochastic trajectories.}
        \label{fig:new_energy_delay_embedding}
    \end{subfigure}
    
    \caption{Local delay embedding at the spatial location $(x_i, y_i) = (\pi, \pi)$ using a delay parameter $\tau = 10\delta t$. The reference manifold is constructed from ten ground-truth trajectories, against which the generated stochastic trajectories are projected for comparison.}
    \label{fig:energy_delay_comparison}
\end{figure}

Each generated trajectory, although initialized from the same initial condition, is unique and does not 'replicate' any training path. To formally assess the plausibility of the generated trajectories, we consider delay-embedding manifolds. Delay embeddings provide a dynamically informative representation of the system, as they encode the distribution of future states conditioned on past states, which is precisely the dependency structure learned by the transition kernel. Figure~\ref{fig:energy_delay_comparison} compares the delay-embedding manifold of the kinetic energy at a fixed spatial location. The kinetic energy,
\[
k = \frac{1}{2}(U^{2} + V^{2}),
\]
provides a compact observable that aggregates the two velocity components. We construct a reference, local manifold from the ground-truth dataset by projecting the kinetic energy into the two-dimensional delay-embedding space \((k_t, k_{t-\tau})_{x_i, y_i}\), with a fixed delay parameter \(\tau\) and at a location $(x_i, y_i)$. The choice of \(\tau\) must balance two constraints: it must be small enough for \((k_t, k_{t-\tau})\) to remain correlated, while being sufficiently large to avoid a trivial identity mapping and reveal the underlying joint distribution. Each generated trajectory is then projected into the same delay-embedding space. As shown in Figure~\ref{fig:energy_delay_comparison}, the generated trajectories largely overlap with the reference manifold, indicating that the learned transition kernel preserves the dynamical structure of the true system and generates trajectories that remain consistent with the underlying attractor.

\begin{figure}[H]
    \centering
    
    \begin{subfigure}[b]{0.36\linewidth}
        \centering
        \includegraphics[width=\linewidth]{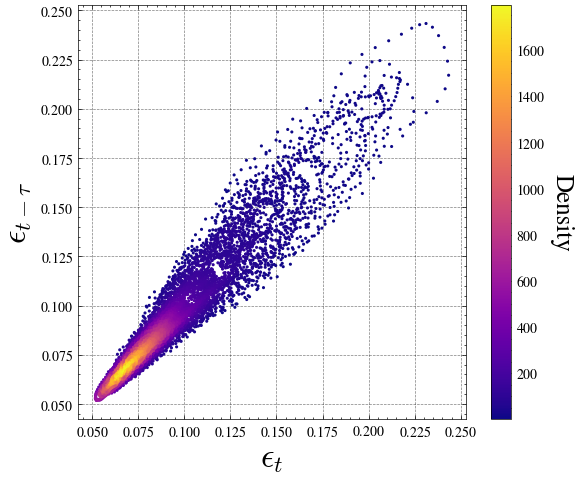}
        \caption{Reference manifold obtained from the dissipation delay embedding.}
        \label{fig:dissipation_delay_embedding}
    \end{subfigure}
    \hfill
    \begin{subfigure}[b]{0.60\linewidth}
        \centering
        \includegraphics[width=\linewidth]{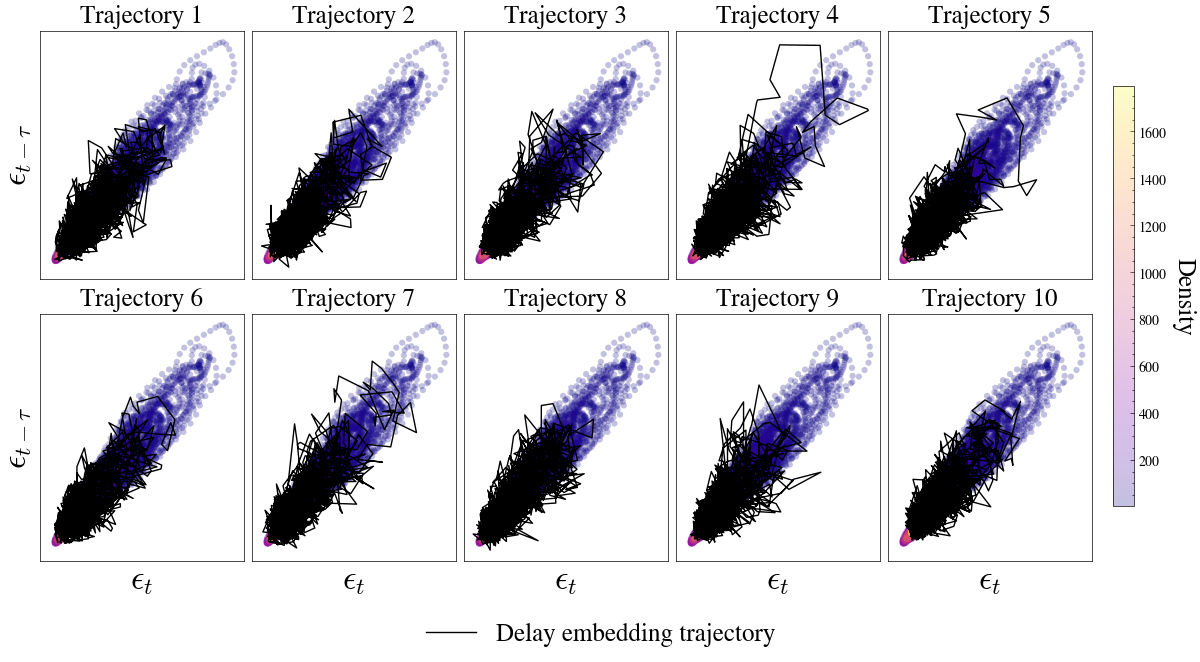}
        \caption{Delay embeddings of the generated stochastic trajectories.}
        \label{fig:new_dissipation_delay_embedding}
    \end{subfigure}
    
    \caption{Global delay embedding using a delay parameter $\tau = 5\delta t$. The reference manifold is constructed from ten ground-truth trajectories, against which the generated stochastic trajectories are projected for comparison.}
    \label{fig:dissipation_delay_comparison}
\end{figure}

To complement the local analysis based on the kinetic energy at a single probe location, we perform the same delay-embedding analysis using the dissipation rate averaged over the full two-dimensional flow field. Unlike the kinetic energy probe, the dissipation provides a global measure of the flow dynamics. The dissipation rate is defined as :
\[
\epsilon(t) = \frac{\nu}{|\Omega|}\int_{\Omega}
\left(
\left(\frac{\partial U}{\partial x}\right)^2
+
\left(\frac{\partial U}{\partial y}\right)^2
+
\left(\frac{\partial V}{\partial x}\right)^2
+
\left(\frac{\partial V}{\partial y}\right)^2
\right)\, d\Omega,
\]
where \(\nu\) is the viscosity and \(\Omega\) denotes the spatial domain. Using this scalar observable, we construct the dissipation delay-embedding manifold in the same way as for the kinetic energy, by projecting the temporal signal into the two-dimensional space \((\epsilon_t,\epsilon_{t-\tau})\) with a fixed delay parameter \(\tau\). Figure~\ref{fig:dissipation_delay_comparison} compares the delay embeddings of the generated trajectories with the reference manifold obtained from the ground-truth system. The strong overlap between the generated trajectories and the reference manifold further supports that the learned transition kernel preserves the global dissipative structure of the true dynamics.

\begin{figure}[H]
\centering

\begin{subfigure}[t]{0.48\linewidth}
\centering
\includegraphics[width=\linewidth]{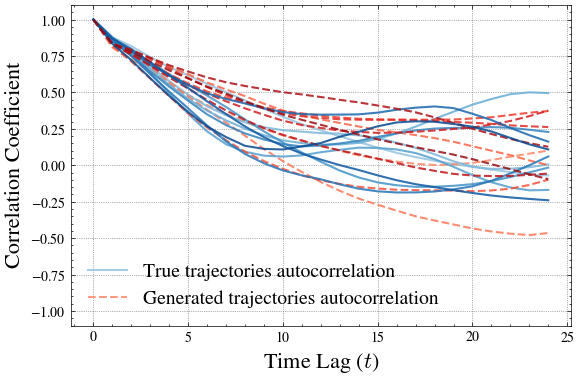}
\caption{Autocorrelation of the ground truth dataset trajectories $U(t)$ and the generated trajectories ($\hat{U}(t)$).}
\label{fig:autocorr_U}
\end{subfigure}
\hfill 
\begin{subfigure}[t]{0.48\linewidth}
\centering
\includegraphics[width=\linewidth]{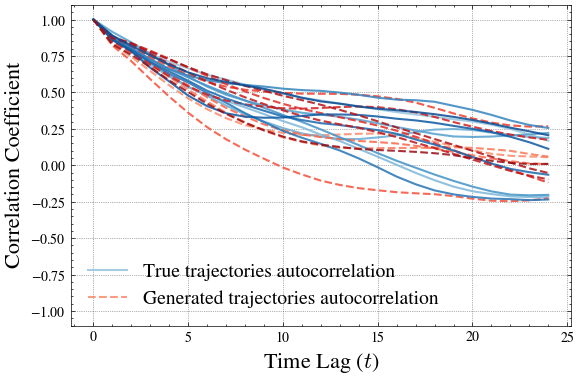}
\caption{Autocorrelation of the ground truth dataset trajectories $V(t)$ and the generated trajectories ($\hat{V}(t)$).}
\label{fig:autocorr_V}
\end{subfigure}
\caption{Autocorrelation functions computed over a horizon of $250\times \delta t$. The Pearson correlation coefficient is evaluated between each trajectory and its time-shifted counterpart at lags separated by $10\times \delta t$.}
\label{fig:autocorr}
\end{figure}

The previous analyses suggest that the learned transition kernel is dynamically consistent with the true system both locally and globally. To further validate this, we analyse the autocorrelation structure of the trajectories in order to quantify how information about the current state is retained as the system evolves. From an information-theoretic perspective, the autocorrelation function provides a measure of memory decay and allows us to assess whether the generated trajectories lose information at a rate comparable to that of the true system. From a dynamical systems perspective, autocorrelation can also reveal characteristic patterns such as periodic or quasi-periodic behaviour, which provide insight into the topology of the underlying attractor. Figure~\ref{fig:autocorr} shows that the learned transition kernel reproduces the autocorrelation structure of the reference system well. In particular, the rate at which autocorrelation decays is consistent between the generated and true trajectories, indicating that the model captures the characteristic memory timescales of the dynamics. 

\begin{figure}[H]
\centering

\begin{subfigure}[t]{0.27\linewidth}
    \centering
    \includegraphics[width=\linewidth]{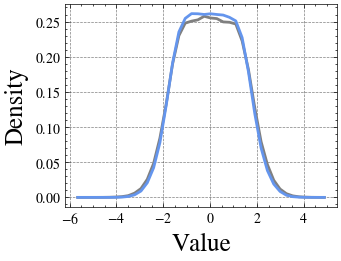}
    \caption{Aggregated density of \(\hat{U}(x,y,t)\) with reference \(U(x,y,t)\) density.}
    \label{fig:U_pdf_agg}
\end{subfigure}
\hfill
\begin{subfigure}[t]{0.70\linewidth}
    \centering
    \includegraphics[width=\linewidth]{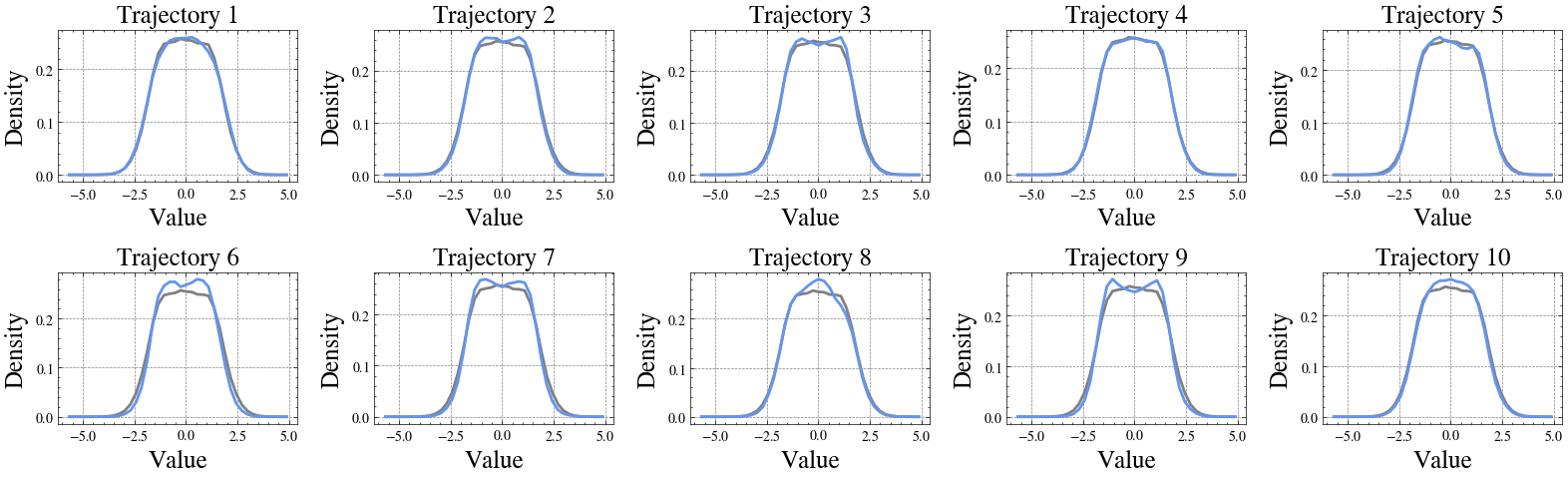}
    \caption{Individual trajectory densities of \(\hat{U_i}(x,y,t)\) with reference \(U(x,y,t)\) density.}
    \label{fig:U_pdf}
\end{subfigure}

\vspace{0.5cm}

\begin{subfigure}[t]{0.27\linewidth}
    \centering
    \includegraphics[width=\linewidth]{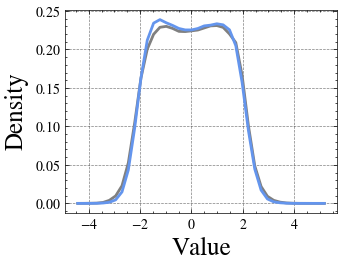}
    \caption{Aggregated density of \(\hat{V}(x,y,t)\) with reference \(V(x,y,t)\) density.}
    \label{fig:V_pdf_agg}
\end{subfigure}
\hfill
\begin{subfigure}[t]{0.70\linewidth}
    \centering
    \includegraphics[width=\linewidth]{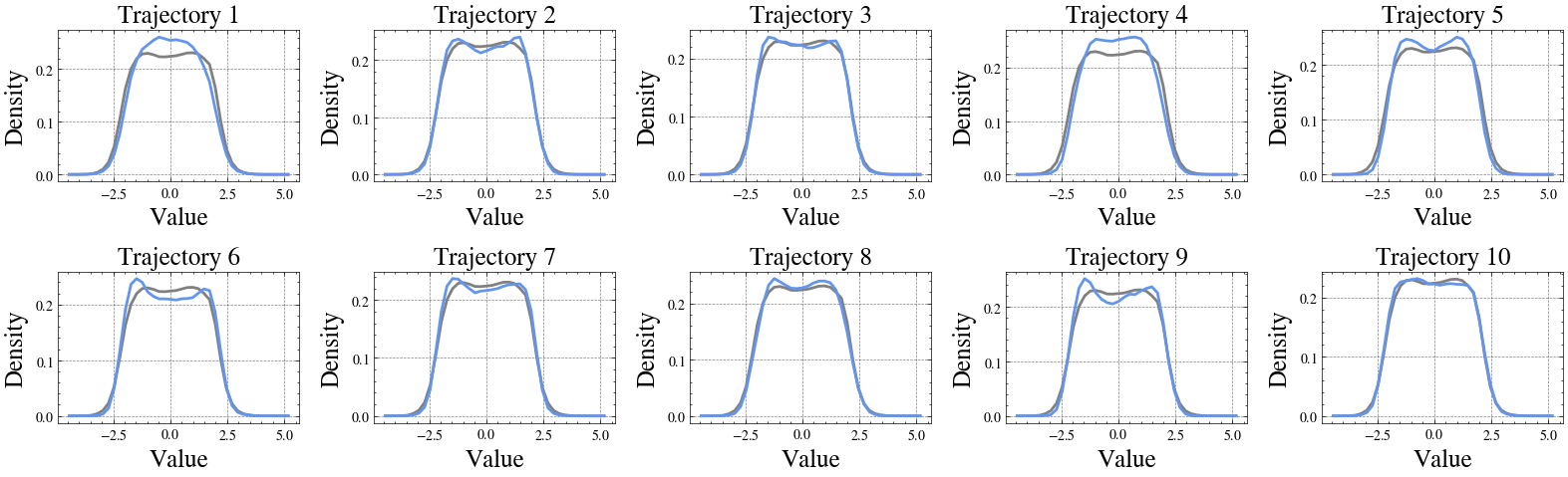}
    \caption{Individual trajectory densities of \(\hat{V_i}(x,y,t)\) with reference \(V(x,y,t)\) density.}
    \label{fig:V_pdf}
\end{subfigure}

\caption{
Comparison of the marginal density functions of the velocity components. The left column shows densities estimated from all generated trajectories aggregated together, whereas the right column displays the densities of the generated trajectories individually. The reference density is estimated by aggregating the ten ground-truth trajectories and serves as an approximation of the invariant distribution that each generated trajectory is expected to reproduce. The top row corresponds to \(U\), and the bottom row to \(V\).
}
\includegraphics[width = 0.4\textwidth]{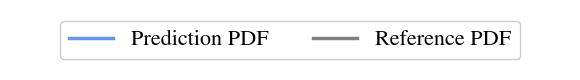}
\label{fig:pdf_comparison}
\end{figure}

To assess whether the generated flow is also statistically consistent with the reference dynamics, we also compare the empirical probability density functions of the generated trajectories with those of the ground-truth system, interpreting the latter as an approximation of the invariant measure of the underlying dynamics. For an ergodic system, long-term trajectories should sample the invariant measure of the attractor. Therefore, if the learned transition kernel faithfully captures the true dynamics, the distribution of generated trajectories should converge toward the same statistical law. Figure~\ref{fig:pdf_comparison} compares the marginal density functions of the velocity components \(U(x,y,t)\) and \(V(x,y,t)\) between the generated trajectories and the reference system. Although individual generated trajectories may exhibit statistical deviations from the reference distribution, as illustrated in Figures~\ref{fig:U_pdf} and \ref{fig:V_pdf}, the distribution obtained by aggregating all generated trajectories closely matches the reference density, as shown in Figures~\ref{fig:U_pdf_agg} and \ref{fig:V_pdf_agg}. This observation suggests that, despite variability at the trajectory level, the stochastic model reproduces the same underlying invariant distribution as the reference system. The close agreement between the aggregated distributions indicates that the generated trajectories accurately capture the system's long-term statistical behaviour.

\begin{figure}[H]
    \centering
    \includegraphics[width=0.4\linewidth]{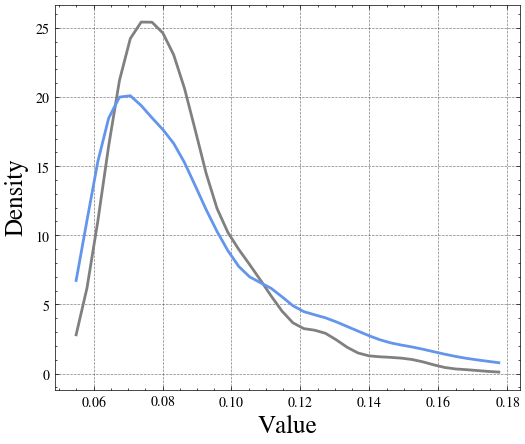}
    \caption{Comparison of the marginal density functions of the dissipation. The reference density is estimated by aggregating the ten ground-truth trajectories and serves as an approximation of the invariant distribution.
}
    \includegraphics[width = 0.4\textwidth]{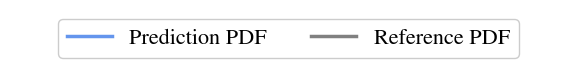}
    \label{fig:dissipation_pdf}
\end{figure}

Similar conclusion can be drawn from different statistics. Figure \ref{fig:dissipation_pdf} compares the density functions of the dissipation signal of our dataset with that of the aggregated, generated ensemble. Although the model's predictions seem less dissipative (also supported by the delay embeddings in figure \ref{fig:dissipation_delay_comparison}), it accurately captures the asymmetry of the system's dissipation statistics or third moment of the probability distribution. The model seem able to explore low-probability regions at the tail of the density function. 

\begin{figure}[H]
    \centering
    \begin{tabular}{c}
        \includegraphics[width=0.5\linewidth]{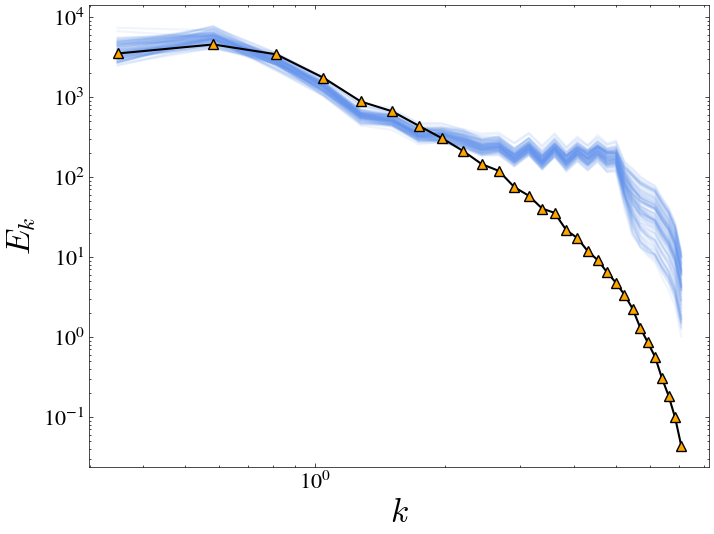} \\
        \includegraphics[width=0.5\linewidth]{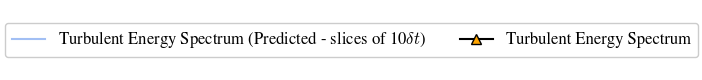}
    \end{tabular}
    \caption{Spectrogram of one generated trajectory averaged over slices of $10\delta t$ compared with the ground-truth reference flow.}
    \label{fig:kolmo_spectrogram}
\end{figure}

Finally, we evaluate the long-rollout stability of the model by checking if the predicted states adhere to a known flow invariant: the Kolmogorov energy cascade. While individual trajectories in data-driven emulators often drift away from this physical reference over time \citep{Thermalizer}, maintaining a consistent scale-energy distribution serves as a crucial sanity check. To quantify this, we compute spectrograms averaged over successive slices of $10\delta t$. As shown in Figure \ref{fig:kolmo_spectrogram}, the generated flow reproduces the energetic contributions across most scales within the inertial subrange. However, energy accumulates artificially at smaller scales (high wavenumbers). This discrepancy between the true and predicted spectrograms emerges at a wavenumber where the corresponding structures account for less than 0.1\% of the total kinetic energy. This discrepancy is likely to originate from decoder reconstruction and is therefore probably a numerical artefact, as it is a recurrent issue associated with convolutional decoders. Although our framework includes an optional spectral regularisation term that corrects the spectrogram faithfully, this arguably degrades spatial reconstruction performance. Hence, it was deactivated for the results presented in this paper. A detailed discussion of this spectral regularisation term and its impact can be found in Appendix \ref{subsec:app-arch_loss}. Crucially, despite this high-wavenumber discrepancy, the simulation remains stable over time and does not blow up.

\subsection{1D - Kuramoto-Sivashinsky}
\label{subsec:results_KS}
Here, the 1D Kuramoto-Sivashinsky dataset is considered. First, we validate performance across both the training and test sets to ensure that the generated trajectories, initialized from the dataset's ground-truth initial conditions, remain consistent with the true system. Figure \ref{fig:10_traj_time_KS} illustrates this evaluation over time at a fixed spatial location $x = \frac{L}{2}$. Visual support is provided in the Appendix \ref{subsec:app-KS}

\begin{figure}[H]
    \centering
    \includegraphics[width=1\linewidth]{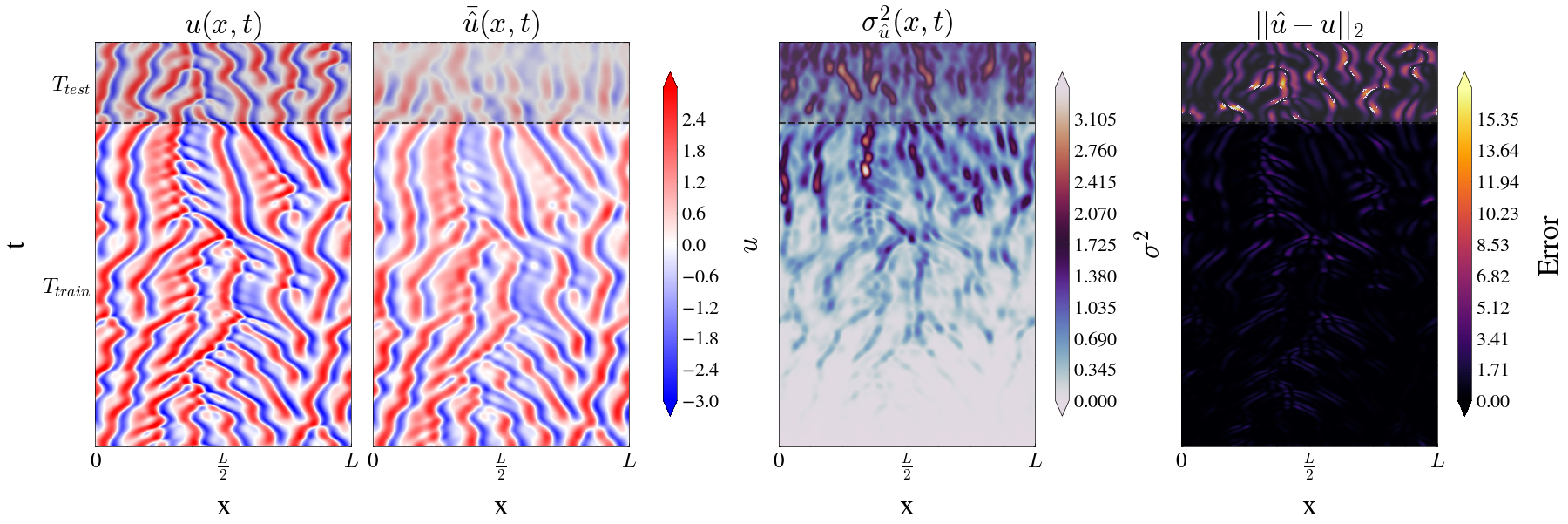}
    \caption{From left to right : Ground truth, corresponding to the last trajectory from the dataset, mean trajectory prediction from the same initial condition, averaged over an ensemble of 60 members, variance of the ensemble and quadratic distance between the ground truth and the mean trajectory. The shaded area highlights the test region.}
    \label{fig:err_UQ_mean}
\end{figure}

\begin{table}[H]

\centering
\caption{Evaluation Metrics across Trajectories}
\label{tab:MSE_PICP}
\begin{tabular}{lccc}
\toprule
 &  MSE &  PICP $1\sigma$ &  PICP $3\sigma$ \\
\midrule
Train Set    & 0.64           & 70\%                    & 96\%                    \\
Test Set     & 2.11          & 54\%                    & 95\%            \\      
\bottomrule
\end{tabular}
\end{table}
Similar conclusions to what was reported in \ref{subsec:exp_kolmo} can be drawn, with performance metrics gathered in  table \ref{tab:MSE_PICP}. Over the entire dataset, the empirical coverage reaches approximately \(66.8\%\) within one standard deviation and \(96\%\) within three standard deviations, indicating that the predictive uncertainty is broadly consistent with Gaussian behaviour, for which the expected coverages are 68.3\% and 99.7\%, respectively. As a sanity check, we observe consistency between the ensemble variance and the deviation of individual trajectories from the ground truth: trajectories tend to separate from one another at approximately the same time they separate from the reference trajectory. This is visually supported by figure \ref{fig:err_UQ_mean} where the increase in variance occurs synchronously with the $L_2$ distance between the mean trajectory and the ground truth. 

This behaviour confirms that individual ensemble members decorrelate from one another at the same rate they diverge from the reference trajectory, while remaining physically plausible realizations.

To evaluate the learned transition kernel beyond simple memorization, we test its generative capability on an unseen, synthetic initial condition outside the training set. We autonomously rollout an ensemble of 10 trajectories over 2000 snapshots, thus once more exceeding the combined training and test window, without a ground-truth reference, similarly to what was reported in \ref{subsec:results_kolmo}. The goal is purely to assess long-term dynamical plausibility.

\begin{figure}[H]
    \centering
    \begin{subfigure}[b]{0.34\linewidth}
        \centering
        \includegraphics[width=\linewidth]{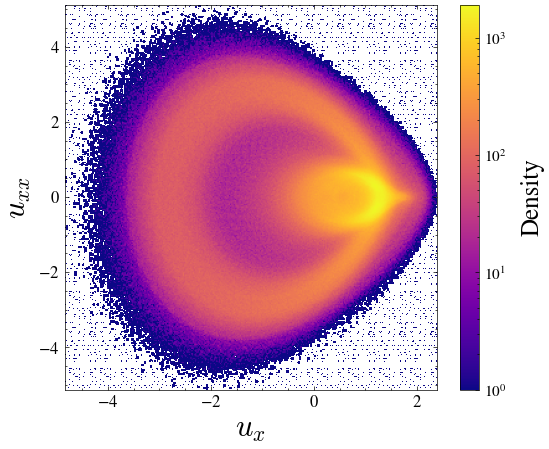}
        \caption{Reference manifold (true data).}
        \label{fig:Ks_ref_manifold}
    \end{subfigure}
    \hfill
    \begin{subfigure}[b]{0.62\linewidth}
        \centering
        \includegraphics[width=\linewidth]{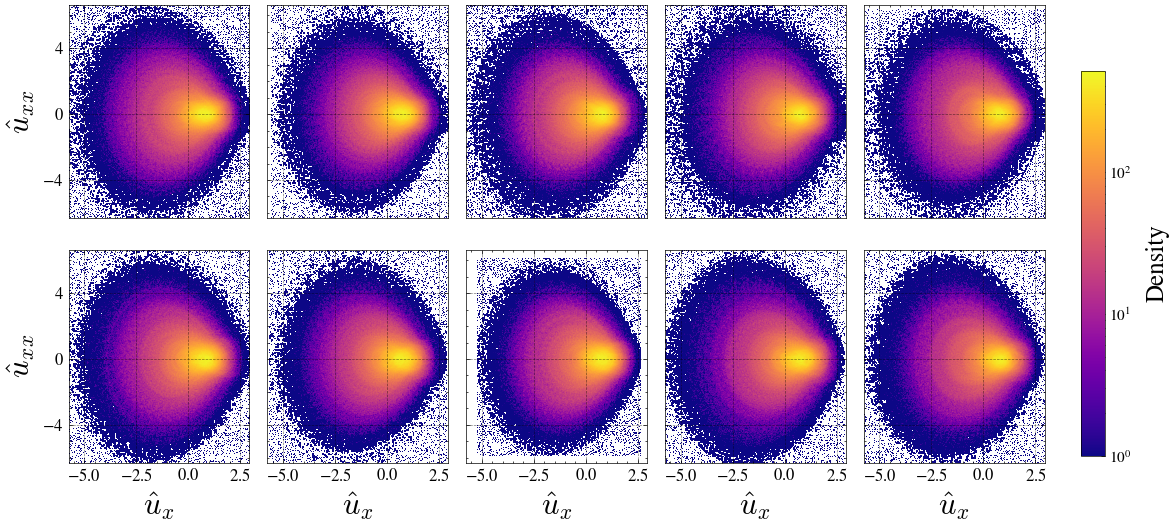}
        \caption{Generated trajectories projection.}
        \label{fig:Ks_pred_manifold}
    \end{subfigure}
    \caption{Low-dimensional manifold in the $(u, u_{xx})$ phase space. The generated stochastic trajectories are projected into the same embedding space for comparison.}
    \label{fig:KS_manifold}
\end{figure}

While each generated trajectory evolves uniquely, they remain strictly bounded by the true system dynamics. Remarkably, when projected into the low-dimensional $(u, u_{xx})$ phase-space, all ten autonomous trajectories accurately capture and respect the topology of the underlying physical manifold (Figure \ref{fig:KS_manifold}).

\begin{figure}[H]
    \centering
    \begin{subfigure}[b]{0.48\linewidth}
        \centering
        \includegraphics[width=\linewidth]{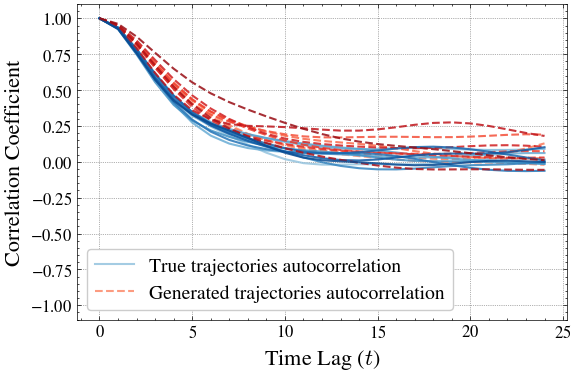}
        \caption{Temporal autocorrelation functions computed over a horizon of $250\delta t$. The Pearson correlation coefficient is evaluated between each trajectory and its time-shifted counterpart at lags separated by $10\delta t$.}
        \label{fig:autocorr_time_KS}
    \end{subfigure}
    \hfill
    \begin{subfigure}[b]{0.48\linewidth}
        \centering
        \includegraphics[width=\linewidth]{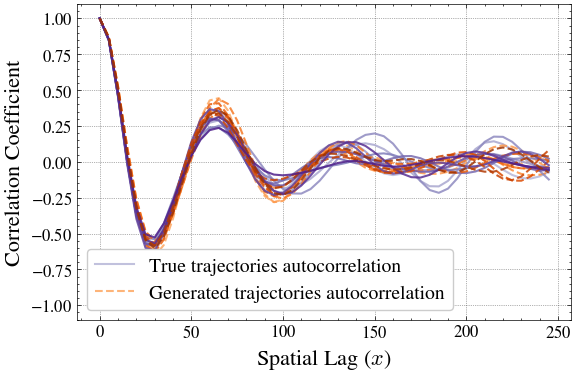}
        \caption{Spatial autocorrelation functions computed over a spatial domain of $250\delta x$. The Pearson correlation coefficient is evaluated between each trajectory and its spatially shifted counterpart at lags separated by $10\delta x$.}
        \label{fig:autocorr_space_KS}
    \end{subfigure}
    \caption{Comparison of temporal (a) and spatial (b) autocorrelation functions between the true and generated trajectories.}
    \label{fig:autocorr_KS}
\end{figure}

Figure~\ref{fig:autocorr_space_KS} demonstrates that the learned transition kernel successfully reproduces both the temporal and spatial autocorrelation structures of the reference system, exhibiting strong agreement with the ground truth. Furthermore, not only is the spatial decorrelation rate preserved, as supported by Figure~\ref{fig:autocorr_space_KS}, but the oscillating behaviour of the autocorrelation function also indicates a spatial pseudo-periodicity. Because the model accurately captures this characteristic feature, it provides strong evidence of agreement between the true and generated attractor topologies.

\begin{figure}[H]
\centering
\includegraphics[width=\linewidth]{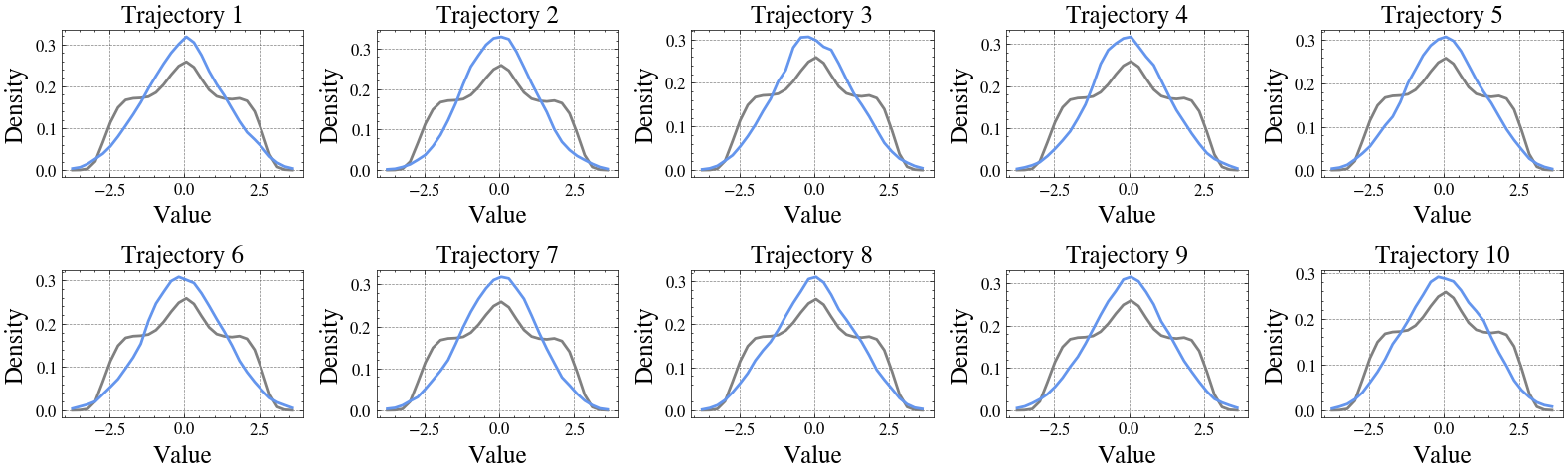}
\caption{Individual trajectory densities of \(\hat{u}\) with reference \(u\) density.}
\label{fig:pdf_comparison_ks}
\includegraphics[width = 0.4\textwidth]{images/results/KS/legend_PDF.png}
\end{figure}

Then, looking at statistical invariants, figure~\ref{fig:pdf_comparison_ks} compares the marginal density functions  between the generated trajectories $\hat{u}(x, t)$ and the reference system $u(x, t)$.
As shown in Figure~\ref{fig:pdf_comparison_ks}, mean, variance, and skewness, are well recovered across all trajectories, confirming that the model captures the main moments faithfully. However, a systematic discrepancy is observed in the fourth moment: the generated PDFs exhibit a consistently lower kurtosis than the reference, manifesting as an over-smooth central peak. Addressing this deficit may require moment matching regularization.

\begin{figure}[H]
    \centering
    \begin{subfigure}[b]{0.47\linewidth}
        \centering
        \includegraphics[width=\linewidth]{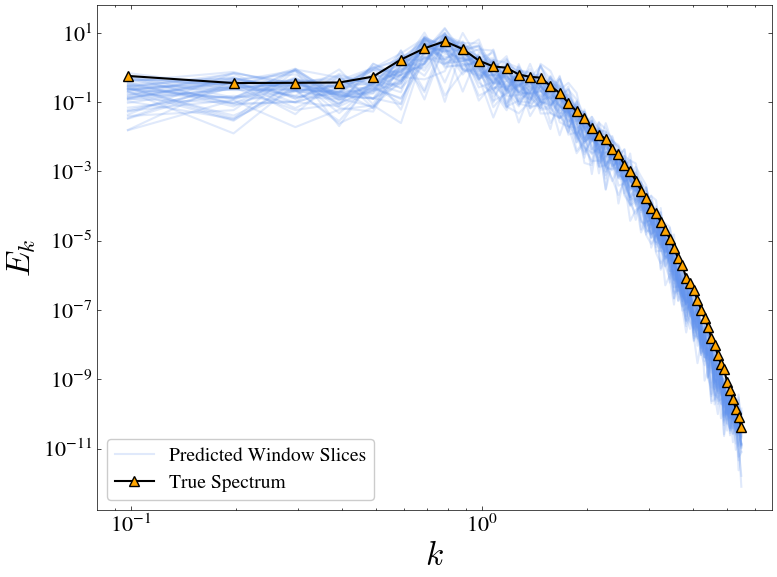}
        \caption{Spatial energy spectrum quantifying the energetic contribution across wavenumbers, averaged over temporal slices of $10\delta t$.}
        \label{fig:spatial_spectrum_KS}
    \end{subfigure}
    \hfill
    \begin{subfigure}[b]{0.47\linewidth}
        \centering
        \includegraphics[width=\linewidth]{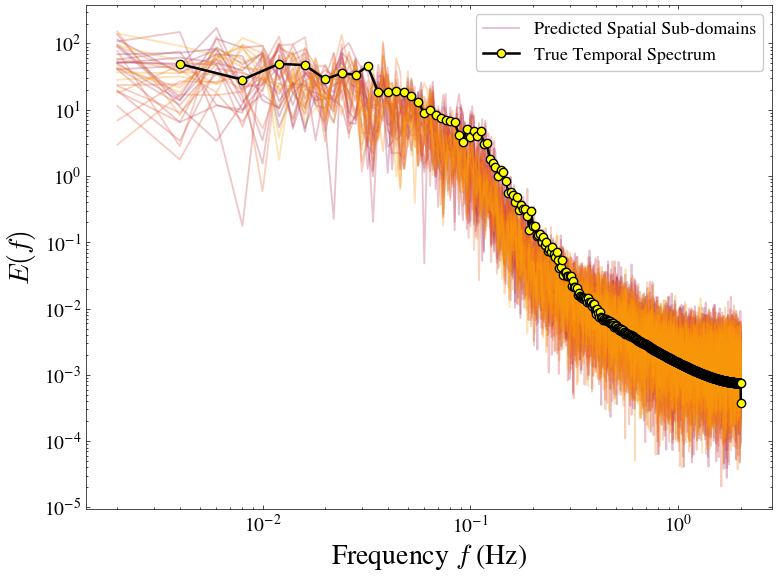}
        \caption{Temporal frequency spectrum quantifying the distribution of dynamic frequencies, averaged over spatial segments of $16\delta x$.}
        \label{fig:temporal_spectrum_KS}
    \end{subfigure}
    \caption{Comparison of the (a) spatial energy spectra and (b) temporal frequency spectra between the true and generated trajectories.}
    \label{fig:spectra_KS}
\end{figure}

Finally, we evaluate model consistency with respect to physical invariants and spectral properties. Specifically, we compute the spatial kinetic energy spectrum—averaged across temporal slices of $10\delta t$, to verify stability over long rollouts and consistency across scale relative contributions. Complementarily, we perform a temporal frequency spectrum analysis across spatial segments of $16\delta x$ to validate the generated trajectories in the Fourier domain. Both spectral comparisons are presented in Figure \ref{fig:spectra_KS}.

Not only do the true and generated spectra closely match—suggesting accurate relative contributions across both scales and frequencies, but Figure~\ref{fig:spatial_spectrum_KS} also demonstrates the stability of this physical invariant as the rollout progresses well beyond the training and testing time windows. Additionally, the artificial energy accumulation at high wavenumbers observed in the Kolmogorov flow test case  (Figure~\ref{fig:kolmo_spectrogram}) does not occur here. Indeed, because the KS experiment is one-dimensional, we utilize a fully connected autoencoder and decoder. This mitigates the convolutional reconstruction artifacts that, in our view, are responsible for the accumulation of energy at very high wavenumbers.

\section{Conclusion}
\label{sec:conclusion}

In this work, we attempt to reframe the role of stochasticity in scientific machine learning, from a tool for uncertainty quantification to a foundational framework for modelling chaotic, non-linear macroscopic physics. Rather than exhausting computational resources to deterministically resolve highly sensitive trajectories through infinite precision, our approach embraces the inherent stochastic nature of chaotic fluid dynamics. By modelling the system statistically, we successfully capture the multi-admissible futures that occur when fluid flows transit between competing attractors.

We established an encode-process-decode framework utilizing an Autoencoder. Within this latent space, a Transformer architecture was leveraged to learn the state-dependent drift and diffusion terms of a discretized Itô Stochastic Differential Equation (SDE).

Given the extreme sensitivity to initial conditions inherent to chaotic systems, evaluating the framework against explicit ground-truth trajectories is fundamentally unrepresentative of its true performance. Instead, we evaluated the models based on their local and global statistical consistency across the flow field in two different test cases. Evaluation via manifold comparison, delay embeddings and autocorrelation plots confirmed that the system preserves long-term correct conditioning of future states by past states thus preserving dynamical consistency. Crucially, our experiment demonstrated stochastic stability over long rollout. From a statistical standpoint, the model seem to recover important statistical invariants. In particular, the first three moments are accurately predicted, even for statistics exhibiting low-probability events such as the dissipation signal. However, 4th moments and statistics exhibiting complicated kurtosis are not recovered at this point.

A critical open question arising from this methodology lies in the intricate mixing of different sources of uncertainty, making it difficult to disentangle aleatoric (inherent physical randomness) from epistemic uncertainty. In our current framework, the learned diffusion term of the SDE acts as a catch-all container absorbing three distinct stochastic drivers. First, it captures the intrinsic chaos and multi-attractor properties of the system, representing the true aleatoric uncertainty of the physical system. Second, it absorbs the subspace projection loss of information, which is an irreducible stochasticity resulting from truncating the state space. And third, it encompasses the model's own epistemic uncertainties, absorbing the approximation errors and structural limitations inherent to the neural network components or lack of training data. This co-existence inherently creates a mathematical conflict: the system's true physical stochasticity, which we aim to learn, is structurally confounded with the model's approximation and architectural uncertainties, which we aim to minimize. Disentangling these mixed uncertainties presents an exciting pathway for future research. Resolving this conflict will improve interpretability and pave the way for more robust probabilistic models across a wider spectrum of chaotic macroscopic systems.
\newpage 

\bibliography{biblio}

@article{POD-Galerkin,
title = {Data-driven POD-Galerkin reduced order model for turbulent flows},
journal = {Journal of Computational Physics},
volume = {416},
pages = {109513},
year = {2020},
issn = {0021-9991},
doi = {https://doi.org/10.1016/j.jcp.2020.109513},
url = {https://www.sciencedirect.com/science/article/pii/S0021999120302874},
author = {Saddam Hijazi and Giovanni Stabile and Andrea Mola and Gianluigi Rozza}
}

@article{CD-ROM,
title = {CD-ROM: Complemented Deep - Reduced order model},
journal = {Computer Methods in Applied Mechanics and Engineering},
volume = {410},
pages = {115985},
year = {2023},
issn = {0045-7825},
doi = {https://doi.org/10.1016/j.cma.2023.115985},
url = {https://www.sciencedirect.com/science/article/pii/S0045782523001081},
author = {Emmanuel Menier and Michele Alessandro Bucci and Mouadh Yagoubi and Lionel Mathelin and Marc Schoenauer}
}

@article{quantized-ROM,
title = {Quantized local reduced-order modeling in time (ql-ROM)},
journal = {Computer Methods in Applied Mechanics and Engineering},
volume = {447},
pages = {118393},
year = {2025},
issn = {0045-7825},
doi = {https://doi.org/10.1016/j.cma.2025.118393},
url = {https://www.sciencedirect.com/science/article/pii/S0045782525006656},
author = {Antonio Colanera and Luca Magri}
}

@article{MZAE,
    author = {Gupta, Priyam and Schmid, Peter and Sipp, Denis and Sayadi, Taraneh and Rigas, Georgios},
    title = {Mori–Zwanzig latent space Koopman closure for nonlinear autoencoder},
    journal = {Proceedings of the Royal Society A: Mathematical, Physical and Engineering Sciences},
    volume = {481},
    number = {2313},
    pages = {20240259},
    year = {2025},
    month = {05},
    issn = {1364-5021},
    doi = {10.1098/rspa.2024.0259},
    url = {https://doi.org/10.1098/rspa.2024.0259},
    eprint = {https://royalsocietypublishing.org/rspa/article-pdf/doi/10.1098/rspa.2024.0259/2822700/rspa.2024.0259.pdf},
}

@article{Koopman,
author = {Brunton, Steven L. and Budi\v{s}i\'{c}, Marko and Kaiser, Eurika and Kutz, J. Nathan},
title = {Modern Koopman Theory for Dynamical Systems},
journal = {SIAM Review},
volume = {64},
number = {2},
pages = {229-340},
year = {2022},
doi = {10.1137/21M1401243},
URL = {  
        https://doi.org/10.1137/21M1401243
},
eprint = {  
        https://doi.org/10.1137/21M1401243

}
}

@article{DMD, 
title={Dynamic mode decomposition of numerical and experimental data}, 
volume={656}, 
DOI={10.1017/S0022112010001217}, 
journal={Journal of Fluid Mechanics}, 
author={Schmid, Peter J.}, 
year={2010}, 
pages={5–28}}

@article{UPdROM,
  author = {Zighed, Ismaël and Thome, Nicolas and Gallinari, Patrick and Sayadi, Taraneh},
  title = {Uncertainty-aware and parametrized dynamic reduced-order model: Application to unsteady flows}, 
  journal = {Phys. Rev. Fluids}, 
  year = {2025},
  publisher = {American Physical Society},
  doi = {10.1103/f6ty-t6gl},
  url = {https://link.aps.org/doi/10.1103/f6ty-t6gl}
}

@misc{ESN,
author = {Özalp, Elise and Nóvoa, Andrea and Magri, Luca},
year = {2025},
month = {08},
pages = {},
title = {Real-time forecasting of chaotic dynamics from sparse data and autoencoders},
doi = {10.48550/arXiv.2508.08729}
}

@misc{LSTM,
author = {Mohan, Arvind and Gaitonde, Datta},
year = {2018},
month = {04},
pages = {},
title = {A Deep Learning based Approach to Reduced Order Modeling for Turbulent Flow Control using LSTM Neural Networks},
doi = {10.48550/arXiv.1804.09269}
}

@article{Transformer,
author = {Solera-Rico, Alberto and Sanmiguel Vila, Carlos and Gómez-López, Miguel and Wang, Yuning and Almashjary, Abdulrahman and Dawson, Scott and Vinuesa, Ricardo},
year = {2024},
month = {02},
pages = {},
title = {$\beta$-Variational autoencoders and transformers for reduced-order modelling of fluid flows},
volume = {15},
journal = {Nature Communications},
doi = {10.1038/s41467-024-45578-4}
}

@misc{NODE,
  author    = {Rojas, Carlos Jos{\'e} Gonzalez and Dengel, Andreas and Dias Ribeiro, Mateus},
  title     = {Reduced-Order Model for Fluid Flows via Neural Ordinary Differential Equations},
  booktitle = {AAAI Spring Symposium: Combining Artificial Intelligence and Machine Learning with Physical Sciences},
  year      = {2021},
  publisher = {AAAI}
}

@article{Kolmogorov, 
title={Leveraging scale separation and stochastic closure for data-driven prediction of chaotic dynamics}, 
volume={7}, 
DOI={10.1017/dce.2026.10045}, 
journal={Data-Centric Engineering}, 
author={Zighed, Ismaël and Thome, Nicolas and Gallinari, Patrick and Sayadi, Taraneh}, 
year={2026}, 
pages={e12}}

@article{SSM,
  author = {Haller, George and Ponsioen, Sten},
  title = {Nonlinear normal modes and spectral submanifolds: existence, uniqueness and use in model reduction},
  journal = {Nonlinear Dynamics},
  volume = {86},
  number = {3},
  pages = {1493--1534},
  year = {2016},
  doi = {10.1007/s11071-016-2974-z},
  url = {https://doi.org/10.1007/s11071-016-2974-z},
  issn = {1573-269X}
}

@article{Chaotic-SSM,
    author = {Liu, Aihui and Axås, Joar and Haller, George},
    title = {Data-driven modeling and forecasting of chaotic dynamics on inertial manifolds constructed as spectral submanifolds},
    journal = {Chaos: An Interdisciplinary Journal of Nonlinear Science},
    volume = {34},
    number = {3},
    pages = {033140},
    year = {2024},
    month = {03},
    issn = {1054-1500},
    doi = {10.1063/5.0179741},
    url = {https://doi.org/10.1063/5.0179741},
    eprint = {https://pubs.aip.org/aip/cha/article-pdf/doi/10.1063/5.0179741/19848841/033140_1_5.0179741.pdf},
}

@article{Mori,
    author = {Mori, Hazime},
    title = {Transport, Collective Motion, and Brownian Motion*)},
    journal = {Progress of Theoretical Physics},
    volume = {4},
    number = {3},
    pages = {423-455},
    year = {1965},
    doi = {10.1143/PTP.33.423},
}

@article{Zwanzig,
  author    = {Robert Zwanzig},
  title     = {Nonlinear generalized Langevin equations},
  journal   = {Journal of Statistical Physics},
  year      = {1973},
  volume    = {9},
  number    = {3},
  pages     = {215--220},
  doi       = {10.1007/BF01008729},
}

@article{SDE,
title = {Numerical solutions of stochastic differential equations – implementation and stability issues},
journal = {Journal of Computational and Applied Mathematics},
volume = {125},
number = {1},
pages = {171-182},
year = {2000},
note = {Numerical Analysis 2000. Vol. VI: Ordinary Differential Equations and Integral Equations},
issn = {0377-0427},
doi = {https://doi.org/10.1016/S0377-0427(00)00467-2},
url = {https://www.sciencedirect.com/science/article/pii/S0377042700004672},
author = {Kevin Burrage and Pamela Burrage and Taketomo Mitsui}
}

@article{StochasticMZ,
title = {Effective Mori-Zwanzig equation for the reduced-order modeling of stochastic systems},
journal = {Discrete and Continuous Dynamical Systems - S},
volume = {15},
number = {4},
pages = {959-982},
year = {2022},
issn = {1937-1632},
doi = {10.3934/dcdss.2021096},
url = {https://www.aimsciences.org/article/id/885db29c-bdff-44a4-a5bd-e33dca8937d9},
author = {Yuanran Zhu and Huan Lei}
}

@misc{Tornado,
title={Adaptive {SDE} Interpolants for Calibrated Probabilistic {PDE} Forecasting},
author={Jorge Mifsut Benet and Armand Kassa{\"\i} Koupa{\"\i} and Ramon Daniel Regueiro-Espino and Nicolas Baskiotis and Patrick Gallinari},
booktitle={AI{\&}PDE: ICLR 2026 Workshop on AI and Partial Differential Equations},
year={2026},
url={https://openreview.net/forum?id=hwy0BvBIQQ}
}

@article{SDE-net,
  title={SDE-Net: Equipping Deep Neural Networks with Uncertainty Estimates},
  author={Kong, Lingkai and Sun, Jimeng and Zhang, Chao},
  journal={Proceedings of Machine Learning Research},
  volume={119},
  year={2020},
  publisher={ML Research Press}
}

@article{Diffusion,
  author    = {Ho, Jonathan and Jain, Ajay and Abbeel, Pieter},
  title     = {Denoising Diffusion Probabilistic Models},
  journal   = {Advances in Neural Information Processing Systems (NeurIPS)},
  volume    = {33},
  pages     = {6840--6851},
  year      = {2020}
}

@article{score_matching,
  title={Score-based generative modeling through stochastic differential equations},
  author={Song, Yang and Sohl-Dickstein, Jascha and Kingma, Diederik P and Kumar, Abhishek and Ermon, Stefano and Poole, Ben},
  journal={arXiv preprint arXiv:2011.13456},
  year={2020}
}

@misc{Neural-SDE,
      title={Neural SDEs as a Unified Approach to Continuous-Domain Sequence Modeling}, 
      author={Macheng Shen and Chen Cheng},
      year={2025},
      eprint={2501.18871},
      archivePrefix={arXiv},
      primaryClass={cs.LG},
      url={https://arxiv.org/abs/2501.18871}, 
}

@article{KolmogorovRe_Nf,
    author = {Platt, N. and Sirovich, L. and Fitzmaurice, N.},
    title = {An investigation of chaotic Kolmogorov flows},
    journal = {Physics of Fluids A: Fluid Dynamics},
    volume = {3},
    number = {4},
    pages = {681-696},
    year = {1991},
    month = {04},
    issn = {0899-8213},
    doi = {10.1063/1.858074},
    url = {https://doi.org/10.1063/1.858074},
    eprint = {https://pubs.aip.org/aip/pof/article-pdf/3/4/681/12301110/681_1_online.pdf},
}

@book{Canuto,
  author       = {Claudio Canuto and Alfio Quarteroni and M. Yousuff Hussaini and Thomas A. Zang},
  title        = {Spectral Methods: Evolution to Complex Geometries and Applications to Fluid Dynamics},
  series       = {Scientific Computation},
  publisher    = {Springer Berlin Heidelberg},
  year         = {2007},
  doi          = {10.1007/978-3-540-30728-0},
  isbn         = {978-3-540-30728-0},
  edition      = {1},
  url          = {https://doi.org/10.1007/978-3-540-30728-0},
  note         = {Published: 30 June 2007, eBook ISBN: 978-3-540-30728-0, Hardcover ISBN: 978-3-540-30727-3, Softcover ISBN: 978-3-642-43395-5},
  pages        = {XXX, 596}
}

@misc{Thermalizer,
author = {Pedersen, Chris and Zanna, Laure and Bruna, Joan},
year = {2025},
month = {03},
pages = {},
title = {Thermalizer: Stable autoregressive neural emulation of spatiotemporal chaos},
doi = {10.48550/arXiv.2503.18731}
}

@misc{AttentionAllYouNeed,
 author = {Vaswani, Ashish and Shazeer, Noam and Parmar, Niki and Uszkoreit, Jakob and Jones, Llion and Gomez, Aidan N and Kaiser, Lukasz and Polosukhin, Illia},
 booktitle = {Advances in Neural Information Processing Systems},
 editor = {I. Guyon and U. Von Luxburg and S. Bengio and H. Wallach and R. Fergus and S. Vishwanathan and R. Garnett},
 pages = {},
 publisher = {Curran Associates, Inc.},
 title = {Attention is All you Need},
 url = {https://proceedings.neurips.cc/paper_files/paper/2017/file/3f5ee243547dee91fbd053c1c4a845aa-Paper.pdf},
 volume = {30},
 year = {2017}
}

@article{AdaptivePathIntegral,
  title={Adaptive path-integral autoencoders: Representation learning and planning for dynamical systems},
  author={Ha, Jung-Su and Park, Young-Jin and Chae, Hyeok-Joo and Park, Soon-Seo and Choi, Han-Lim},
  journal={Advances in Neural Information Processing Systems},
  volume={31},
  year={2018}
}

@article{SDEMatching,
  title={Sde matching: Scalable and simulation-free training of latent stochastic differential equations},
  author={Bartosh, Grigory and Vetrov, Dmitry and Naesseth, Christian A},
  journal={arXiv preprint arXiv:2502.02472},
  year={2025}
}

@article{CourseNair,
  title={State estimation of a physical system with unknown governing equations},
  author={Course, Kevin and Nair, Prasanth B},
  journal={Nature},
  volume={622},
  number={7982},
  pages={261--267},
  year={2023},
  publisher={Nature Publishing Group UK London}
}

@article{DDStoROM,
  title={Data-driven stochastic reduced-order modeling of parametrized dynamical systems},
  author={Ilersich, Andrew F and Course, Kevin and Nair, Prasanth B},
  journal={arXiv preprint arXiv:2601.10690},
  year={2026}
}

@article{FM,
  title={Flow matching for generative modeling},
  author={Lipman, Yaron and Chen, Ricky TQ and Ben-Hamu, Heli and Nickel, Maximilian and Le, Matt},
  journal={arXiv preprint arXiv:2210.02747},
  year={2022}
}
\newpage 

\section{Appendix}
\label{Appendix}

\subsection{Kuramoto-Sivashinsky - complementary discussions}
\label{subsec:app-KS}

\begin{figure}[H]
    \centering
    \includegraphics[width=0.6\linewidth]{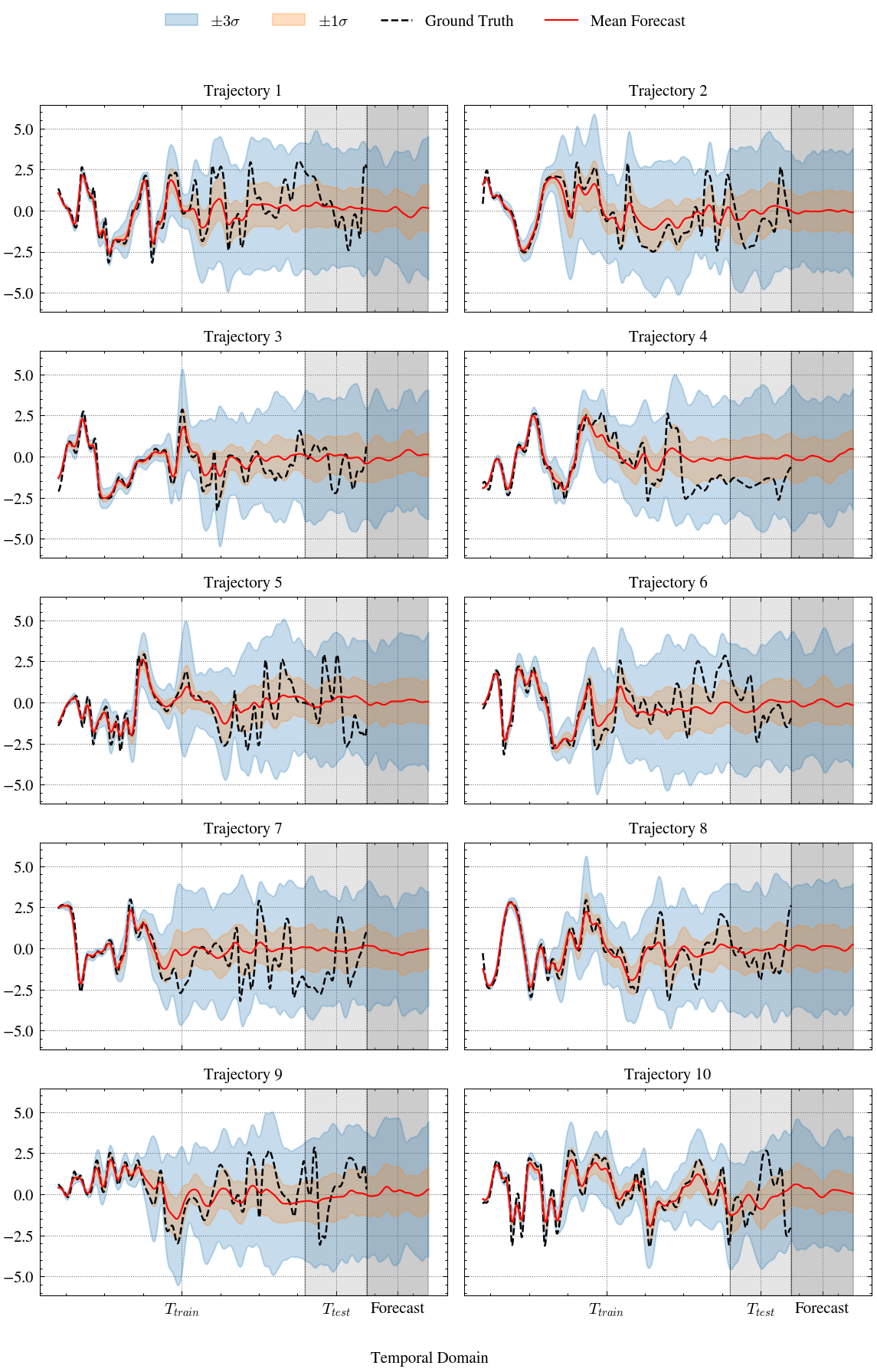}
    \caption{Reconstruction of the field $u(x, t)$ at the spatial location $x = \frac{L}{2}$ over an autoregressive rollout horizon of $T = 1200\,\delta t$. For each initial condition, an ensemble of 60 stochastic trajectories is generated by the model. The solid line denotes the ensemble mean, while the shaded regions represent the intervals corresponding to one and three standard deviations around the mean, illustrating the growth of uncertainty over time.}
    \label{fig:10_traj_time_KS}
\end{figure}

To complement the discussions and statements made in \ref{subsec:exp_ks} we provide here visual support in the Kurmaoto-Sivashinsky benchmark. Figure \ref{fig:10_traj_time_KS}, illustrates how the model rolls out from the initial conditions taken from the training dataset. Following a similar procedure than in \ref{subsec:results_kolmo}, the rollout window encompasses $1200 \delta t$ thus exceeding training and testing window. For this test-case as well, we notice good accordance between the mean trajectory and the true one until a certain time after which, trajectories diverge. This divergence however preserves good statistical consistency as the ground truth remains bounded within a three standard deviation range.

\begin{figure}[H]
    \centering
    \includegraphics[width=1\linewidth]{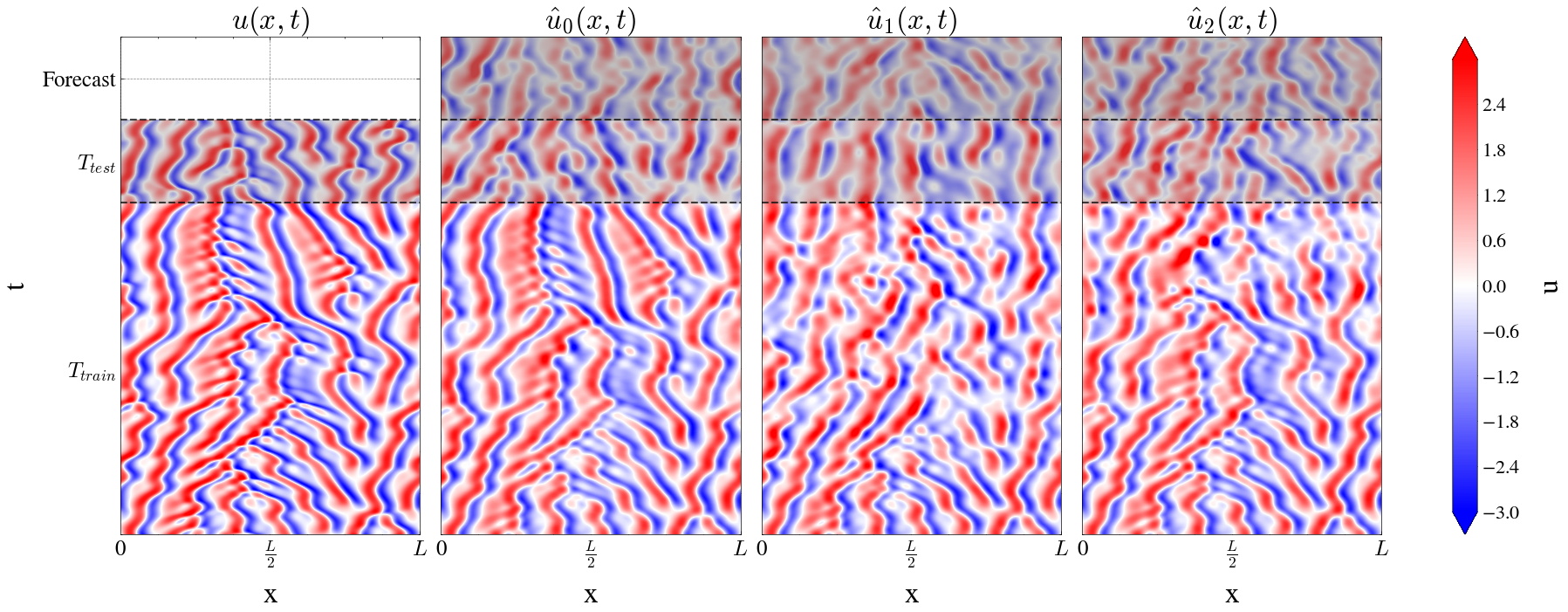}
    \caption{Comparison between a reference trajectory $u(x,t)$ and three stochastic realizations $\hat{u}_i(x,t)_{i=1,2,3}$ initialized from the same initial condition. Snapshots show selected profiles across the training, test, and extended forecast windows ($t \in [1000, 1200] \times \delta t$).}
    \label{fig:10_traj_space_ks}
\end{figure}

In the KS test case as well, individual trajectories sampled from the transition kernel throughout long rollout diverge from their DNS counterparts and from each other, but remain physically plausible, as supported by the figure \ref{fig:10_traj_space_ks}.

\subsection{Model architecture, hyperparameters and loss functions}
\label{subsec:app-arch_loss}

Depending on the configuration, typically if the input data is 1D or 2D, the autoencoder has an adaptative architecture. In the 2D settings, developed for a Kolmogorov flow at $Re=90$ and $n_f =4$
, the autoencoder $(\mathcal{E,D})$ has the following architecture : 

\begin{table}[H]
\centering
\caption{Architecture of the convolutional autoencoder (CNN). The latent dimension is denoted by $d$.}
\label{tab:arch_CNN}
\small
\begin{tabular}{llll}
\toprule
\textbf{Stage} & \textbf{Layer} & \textbf{Output Shape} & \textbf{Details} \\
\midrule

\multicolumn{4}{c}{\textbf{Encoder} : $\mathcal{E}$} \\
\midrule

Input &
-- &
$2 \times 64 \times 64$ &
$\varphi$ \\

Conv1 &
Conv2D + SiLU &
$8 \times 32 \times 32$ &
Kernel $3\times3$, stride $2$\\

Conv2 &
Conv2D + SiLU &
$16 \times 16 \times 16$ &
Kernel $3\times3$, stride $2$\\

Conv3 &
Conv2D + SiLU &
$32 \times 8 \times 8$ &
Kernel $3\times3$, stride $2$\\

Conv4 &
Conv2D &
$64 \times 4 \times 4$ &
Kernel $3\times3$, stride $2$\\

Flatten &
-- &
1024 &
$64 \times 4 \times 4$ \\

Latent projection &
Linear &
$d$ & Dense connections \\

\midrule

\multicolumn{4}{c}{\textbf{Decoder} : $\mathcal{D}$} \\
\midrule

FC Decoder &
Linear &
1024 &
$d \rightarrow 64 \times 4 \times 4$ \\

Reshape &
-- &
$64 \times 4 \times 4$ &
-- \\

Up1 &
Upsample + Conv2D + SiLU &
$32 \times 8 \times 8$ &
Scale factor $2$, kernel $3\times3$ \\

Up2 &
Upsample + Conv2D + SiLU &
$16 \times 16 \times 16$ &
Scale factor $2$, kernel $3\times3$ \\

Up3 &
Upsample + Conv2D + SiLU &
$8 \times 32 \times 32$ &
Scale factor $2$, kernel $3\times3$ \\

Up4 &
Upsample + Conv2D &
$2 \times 64 \times 64$ &
Scale factor $2$, kernel $3\times3$ \\

Output &
Reconstruction &
$2 \times 64 \times 64$ &
$\hat{\varphi}$ \\

\bottomrule
\end{tabular}
\end{table}

The latent transformer $\mathcal{T}$ has the following architecture : 

\begin{table}[H]
\centering
\caption{Transformer model hyperparameters.}
\label{tab:arch_transfo}
\begin{tabular}{lc}
\toprule
\textbf{Hyperparameter} & \textbf{Value} \\
\midrule
Input sequence length ($\tau$) & 20 \\
Transformer dimension & $d$ \\
Attention heads & 8 \\
Attention blocks & 4 \\
Minimum log-diffusion & $-6$ \\
Maximum log-diffusion & $2$ \\
\bottomrule
\end{tabular}
\end{table}

The total training loss function is defined as:
\begin{align}
\label{eq:loss_total}
\mathcal{L}_{\text{total}} &= \mathbb{E}_{z \sim q(z \mid \varphi)} \left[ \left\| \varphi - \mathcal{D}(z) \right\|^2 \right] + \nonumber \\
&\quad + \alpha \mathbb{E}_{(z_{t+1},\,\hat{y}_t)\sim \mathbb{Q}_\theta} \left[ -\log p_\theta(z_{t+1}\mid \hat{y}_t) \right] \nonumber \\
&\quad + \beta \mathbb{E}_{(z_{t+1},\,\hat{y}_t)\sim \mathbb{Q}_\theta} \left[ \big\| \varphi_{t+1} - \mathcal{D}(\hat{z}_{t+1}) \big\|^2 \right] + \gamma E.
\end{align}
In practice, if the Autoencoder is not variational, the reconstruction loss $\mathbb{E}_{z \sim q(z \mid \varphi)} \left[ \left\| \varphi - \mathcal{D}(z) \right\|^2 \right]$ is simply the mean squared error between the reconstructed states $\hat{\varphi}$ and the true states $\varphi$. 

The transition model $\mathcal{T} : \hat{y}_t \mapsto (f_\theta, g_\theta)$ parametrizes the latent dynamics such that:
\begin{equation}
\hat{z}_{t+1} = \hat{z_t} + f_\theta(\hat{y}_t) + g_\theta(\hat{y}_t) \circ \xi, \quad \text{with } \xi \sim \mathcal{N}(0, \mathbf{I}_{d}),
\end{equation}
yielding the transition negative log-likelihood:
\begin{equation}
-\log p_\theta(z_{t+1} \mid \hat{y}_t) = \frac{1}{2}\sum_{i=1}^{d} \left[
\frac{\bigl(z_{t+1}^{(i)} - \hat{z}_t^{(i)} - f_i(\hat{y_t})\bigr)^2}{g_i^2(\hat{y_t})}
+ \log g_i^2(\hat{y_t})
\right].
\end{equation}
The input context history vector $\hat{y}_t$ spanning a temporal window $\tau$ adapts dynamically during the autoregressive rollout. Let $h \geq 1$ denote the current rollout prediction step. The history vector is structured as:
\begin{equation}
\hat{y}_t = \begin{bmatrix} w_t & w_{t-1} & \dots & w_{t-\tau+1} \end{bmatrix}^T,
\end{equation}
where the components $w_{t-i}$ are sampled from either the ground-truth latents or the model's own prior predictions according to:
\begin{equation}
w_{t-i} = \begin{cases} 
\hat{z}_{t-i} & \text{if } i < h, \\
z_{t-i} & \text{if } i \ge h.
\end{cases}
\end{equation}

In contrast to the formulation presented in Eq.~\ref{eq:loss_dynROM}, the objective function in Eq.~\ref{eq:loss_total} introduces a deterministic trajectory regularization term designed to encourage the model to fit training trajectories. This includes a trajectory reconstruction loss,
\begin{equation}
\mathcal{L}_{\text{recon}} = \mathbb{E}_{(z_{t+1},\,\hat{y}_t)\sim \mathbb{Q}_\theta} \left[ \big\| \varphi_{t+1} - \mathcal{D}(\hat{z}_{t+1}) \big\|^2 \right],
\end{equation}
alongside an energy regularization term $E$. The energy loss is computed as the time-dependent energy signal averaged over the spatial domain:
\begin{equation}
E = \mathbb{E}_{(z_{t+1},\,\hat{y}_t)\sim \mathbb{Q}_\theta} \left[ \big\| \langle \hat{U}^{2} + \hat{V}^{2} \rangle_{(x,y)} - \langle U^{2} + V^{2} \rangle_{(x,y)} \big\|^2 \right],
\end{equation}
where $\langle \cdot \rangle_{x}$ denotes the spatial mean operator over the domain $x$. Physically, the energy term enforces that the total energy across the entire velocity field remains bounded within a valid range.

To evaluate the model over extended horizons, we sample $\hat{y}_t$ sequentially $H$ times from the model-induced rollout distribution $\mathbb{Q}_\theta$. Crucially, by setting $H > \tau$, the model is explicitly regularized to generate stable predictions from autoregressive input contexts sampled entirely from its own rollout distribution $\mathbb{Q}_\theta$. The total loss function is computed and averaged over $H$ rollout steps. \\

In the 1D setting, developed for the Kuramoto-Sivashinsky test case, the autoencoder $(\mathcal{E,D})$ has the following architecture :

\begin{table}[H]
\centering
\caption{Architecture of the 1D Autoencoder (AE). It is a fully connected network. The latent dimension is denoted by $d$.}
\label{tab:cvae_architecture_fc}
\small
\begin{tabular}{llll}
\toprule
\textbf{Stage} & \textbf{Layer} & \textbf{Output Shape} & \textbf{Details} \\
\midrule

\multicolumn{4}{c}{\textbf{Encoder} : $\mathcal{E}$} \\
\midrule

Input &
-- &
$512$ &
$\varphi$ \\

FC1 &
Linear + SiLU &
256 &
$512 \rightarrow 256$ \\

FC2 &
Linear + SiLU &
128 &
$256 \rightarrow 128$ \\

FC3 &
Linear  &
d &
$128 \rightarrow d$ \\

\midrule

\multicolumn{4}{c}{\textbf{Decoder} : $\mathcal{D}$} \\
\midrule

FC1 &
Linear + SiLU &
128 &
$d \rightarrow 128$ \\

FC2 &
Linear + SiLU &
256 &
$128 \rightarrow 256$ \\

FC3 &
Linear  &
512 &
$256 \rightarrow \hat{\varphi}$ \\

\bottomrule
\end{tabular}
\end{table}

The hyper-parameter selection applies to both test-case and are summarized in table \ref{tab:hyper-parameter-selection}

\begin{table}[H]
\centering
\caption{Training and model hyperparameters}
\label{tab:hyper-parameter-selection}
\begin{tabular}{lc}
\toprule
\textbf{Hyperparameter} & \textbf{Value} \\
\midrule
Latent dimension ($d$) & 64 \\
Prediction horizon ($H$) & 40 \\
Context sequence length ($\tau$) & 20 \\
Likelihood weight ($\alpha$) & $0.01d^{-1}$ \\
Reconstruction weight ($\beta$) & $0.01$ \\
Energy regularization ($\gamma$) & $0.01$ \\
\bottomrule
\end{tabular}
\end{table}

Finally, it is possible to toggle an ultimate regularization term in the loss, to enforce spectral consistency. The proposed added objective incorporates a kinetic energy spectrogram loss $\mathcal{L}_{E_k}$. This term penalizes discrepancies in the energy cascade spectrogram computed in Fourier space. Because the energy spectrogram acts as a flow invariant, enforcing this constraint encourages the reconstructed velocity fields to preserve the correct multi-scale energy distribution. 

Without this spectral regularization, optimizing solely on an Euclidean distance in physical space leads to reconstructed spectrograms that accumulate unphysical energy in small-scale structures. This occurs because small-scale discrepancies account for only a negligible fraction of the total kinetic energy, meaning a physical-space loss cannot sufficiently penalize high-wavenumber spectral errors. 

To enforce structural fidelity across the entire energy cascade, we define a spectral loss term computed in the Fourier domain. Let $\mathbf{u}' = \mathbf{u} - \langle \mathbf{u} \rangle_{(x,y)}$ represent the fluctuating components of a velocity field. Its 2D Discrete Fourier Transform is denoted by $\tilde{\mathbf{u}}(\mathbf{k}) = (\tilde{u}(\mathbf{k}), \tilde{v}(\mathbf{k}))^T$, where $\mathbf{k} = (k_x, k_y)$ is the wavenumber vector. We extract the radial kinetic energy spectrum by slicing the spectral energy density over $N_k$ wavenumber batches $B_k$:
\begin{equation}
E(k; \mathbf{u}') = \sum_{\mathbf{k} \in B_k} \frac{1}{2} \left( \left| \tilde{u}(\mathbf{k}) \right|^2 + \left| \tilde{v}(\mathbf{k}) \right|^2 \right),
\end{equation}
where $B_k = \{ \mathbf{k} \in \mathbb{N}^2 \mid k \le \|\mathbf{k}\|_2 < k + 1 \}$. 

\begin{figure}[H]
    \centering
    \begin{minipage}{0.48\textwidth}
        \begin{subfigure}{\textwidth}
            \includegraphics[width=1\linewidth]{images/results/Kolmogorov/Spectogram.png}
            \caption{Comparison between one generated trajectory spectrogram averaged over slices of $10\delta t$ and the true spectrogram without spectral regularization in the total loss.}
        \end{subfigure}
    \end{minipage}
    \hfill
    \begin{minipage}{0.48\textwidth}
        \centering
        \begin{subfigure}{\textwidth}
            \includegraphics[width=1\linewidth]{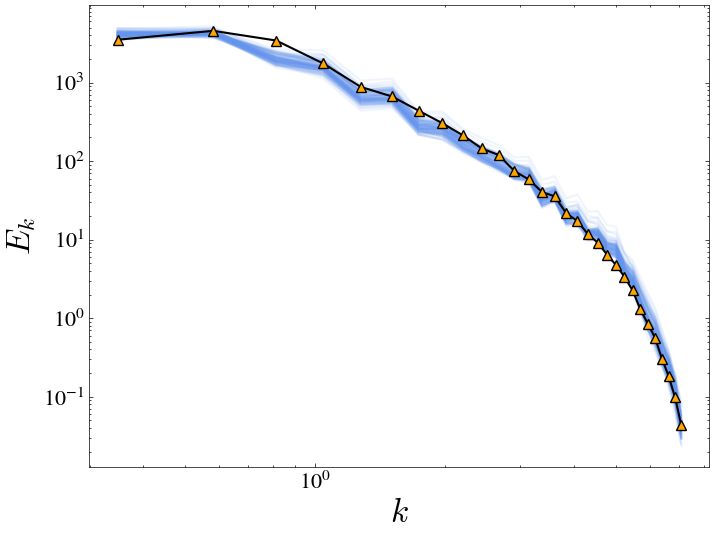}
            \caption{Comparison between one generated trajectory spectrogram averaged over slices of $10\delta t$ and the true spectrogram with spectral regularization in the total loss.}
        \end{subfigure}
    \end{minipage}
    \includegraphics[width = 0.6\linewidth]{images/results/Kolmogorov/spectrogram_legend.png}
    \caption{Impact of spectral regularization on the trajectories spectrograms throughout long-rollouts.}
    \label{fig:spectrogram_comparison}
\end{figure}

The kinetic energy spectrogram loss $\mathcal{L}_{E_k}$ is formalized as the mean squared error in log-space across all $N_k$ resolved discrete radial wavenumbers between the predicted and true spectra, $\hat{E}(k)$ and $E(k)$ respectively. It is given by equation \ref{eq:loss_spectra}. 

\begin{equation}
\label{eq:loss_spectra}
\mathcal{L}_{E_k} = \frac{1}{N_k} \sum_{k=1}^{N_k} \left( \log\left[\hat{E}(k) + \epsilon\right] - \log\left[{E}(k) + \epsilon\right] \right)^2,
\end{equation}
where $\epsilon \ll 1$ ensures numerical stability. The current model applied this regularization for the Kolmogorov test-case and has simply been added to the Autoencoder loss. \\
Figure \ref{fig:spectrogram_comparison} compares the impact of that loss term upon the generated trajectories spectrograms. 

The cost on reconstruction metrics are quantified in table \ref{tab:MSE_PICP_LEK}.

\begin{table}[H]
\centering
\caption{Evaluation Metrics across Trajectories with and without the spectral regularization $\mathcal{L}_{E_k}$ during training.}
\label{tab:MSE_PICP_LEK}
\begin{tabular}{llccc}
\toprule
Configuration & & Mean MSE & Mean PICP $1\sigma$ & Mean PICP $3\sigma$ \\
\midrule
\multirow{2}{*}{Without Spectral Regularization} 
 & Train Set & \textbf{0.64} & \textbf{70}\% & \textbf{96}\% \\
 & Test Set  & \textbf{2.11} & \textbf{54}\% & \textbf{95}\% \\
\midrule
\multirow{2}{*}{With Spectral Regularization} 
 & Train Set & 1.47 & 55\% & 88\% \\
 & Test Set  & 2.12 & \textbf{54}\% & 91\% \\
\bottomrule
\end{tabular}
\end{table}
The addition of the spectral regularization $\mathcal{L}_{E_k}$ in the total loss function slightly degrades reconstruction performances. We suggest that applying it is case-specific, as the associated costs and benefits require a careful compromise.

\subsection{Sensitivity to trajectories }
\label{Appendix:ablation}

Both the KS and Kolmogorov models are initially trained on $10$ distinct initial conditions, each integrated over an $800\delta t$ trajectory generated by a numerical solver. A critical question, however, is how the diversity of training trajectories influences the learned transition kernel's ability to sample a broad, physically sound distribution of futures conditioned on past states. 

The intuition behind this reasoning is best framed through its asymptotic limits. When trained on a single trajectory, the transition kernel lacks exposure to phase-space variability; it becomes overconfident and mode-collapsed, attempting to match predictions to the single sequence observed during training. Conversely, with an infinite ensemble of trajectories, the universal approximation property guarantees that the model perfectly learns the true transition kernel across all valid state sequences permitted by the governing dynamics. Practical interest thus lies in identifying the threshold of training diversity at which the model learns that a given context admits multiple valid future trajectories, while ensuring those futures remain strictly constrained to the system's invariant geometric structure. 

\begin{figure}[htbp]
    \centering
    \begin{subfigure}[b]{0.60\textwidth}
        \centering
        \includegraphics[width=\linewidth]{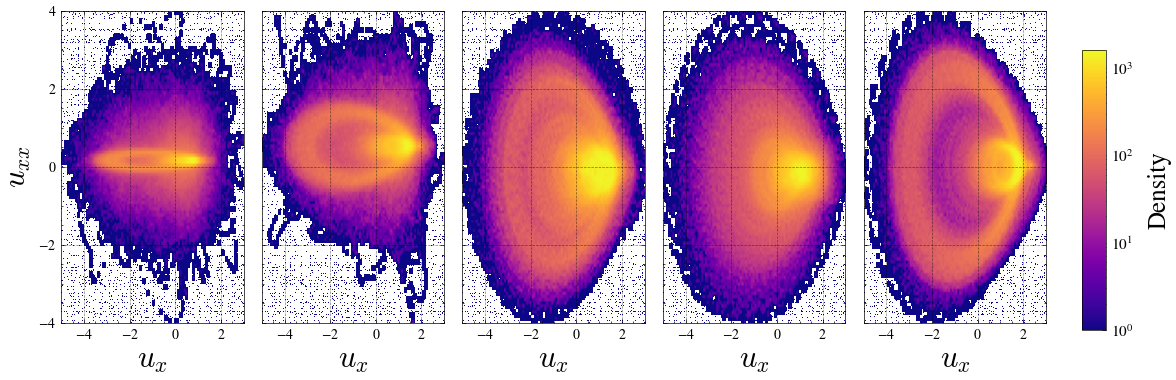}
        \caption{Reconstructed $(u_x, u_{xx})$ phase-space manifolds across training setups. From left to right: model predictions using $1$, $2$, $5$, $10$ training trajectories, DNS generated trajectory.}
        \label{fig:ks_manifold_ablation}
    \end{subfigure}
    \hfill
    \begin{subfigure}[b]{0.38\textwidth}
        \centering
        \includegraphics[width=\linewidth]{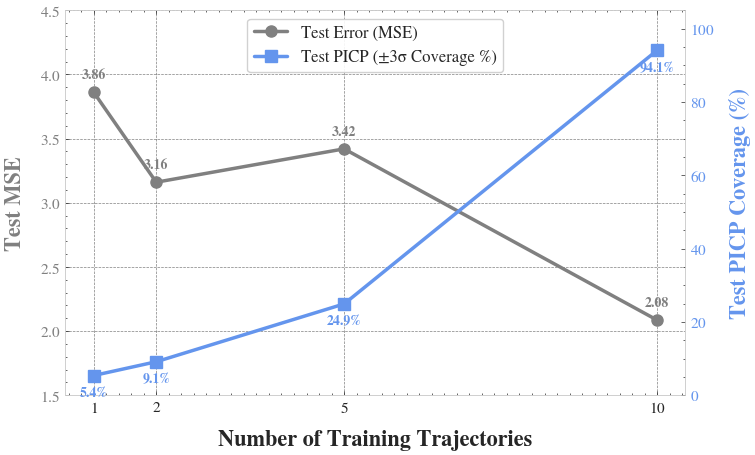}
        \caption{Test MSE and $\text{PICP}_{\pm 3\sigma}$ coverage metrics.}
        \label{fig:ks_metrics_ablation}
    \end{subfigure}
    
    \caption{Impact of training trajectory diversity ($1$, $2$, $5$, and $10$ trajectories) on generative performance for the KS test case. Increasing trajectory diversity improves phase-space manifold reconstruction (\subref{fig:ks_manifold_ablation}) and significantly enhances prediction interval coverage (\subref{fig:ks_metrics_ablation}).}
    \label{fig:traj_number_impact}
\end{figure}

To visualize this structure, we examine the KS system's phase-space manifold. Specifically, we conduct an ablation study by training the model on $1$, $2$, $4$, and $10$ (the presented framework) trajectories, evaluating how accurately and completely the $(u_x, u_{xx})$ manifold is reconstructed during pure generative rollouts initialized from an unseen initial condition. The results are compiled in Figure \ref{fig:traj_number_impact} along with the mean squared error and $\text{PICP}_{\pm 3\sigma}$. 

At first glance, Figure~\ref{fig:ks_manifold_ablation} suggests that training on $5$ trajectories, half the baseline dataset, is sufficient to capture the geometry of the phase-space manifold, even appearing slightly closer to the ground truth than the $10$-trajectory configuration. However, Figure~\ref{fig:ks_metrics_ablation} reveals why $10$ trajectories are necessary for robust generative forecasting. While $5$ trajectories capture the invariant structure, the model severely underestimates uncertainty, yielding a Prediction Interval Coverage Probability ($\text{PICP}_{\pm 3\sigma}$) of only $25\%$. In contrast, $10$ trajectories achieve a coverage of $95\%$, demonstrating that greater dataset diversity is required to reliably encompass the full spectrum of admissible future dynamics.

\subsection{Girsanov KL divergence derivation}
\label{appendix:Girsanov_SOC}

This section details the derivation of the Kullback-Leibler divergence between the two probability measure $\mathbb{P,Q}$ expressed in terms of the Radon–Nikodym derivative, leveraging Stochastic Optimal Control theory and the Girsanov theorem. 

Left to its natural dynamics, the prior SDE generates trajectories that wander blindly and almost never align with the discrete observations $\Phi$. To resolve this, \citep{AdaptivePathIntegral} introduce a control variable $u(z_t, t)$ that acts as a continuous steering force, actively guiding the probability mass of the surrogate measure $\mathbb{Q}_\xi$ toward trajectories that fit the data. 

They parametrize the surrogate drift $h_\xi$ by perturbing the prior drift $f_\theta$ via this control, scaled by the diffusion coefficient:
\begin{equation}
    h_\xi(z_t, t, \Phi) = f_\theta(z_t, t) + g_\theta(z_t, t)u(z_t, t, \Phi).
\end{equation}
Consequently, the control effort required to match the surrogate drift is:
\begin{equation}
\label{eq:control_var}
    u(z_t, t) = g_\theta^{-1}(z_t, t) \big[ h_\xi(z_t, t, \Phi) - f_\theta(z_t, t) \big].
\end{equation}
The log of the Radon–Nikodym derivative reads :
\begin{equation}
    \log \frac{d\mathbb{Q}_\xi}{d\mathbb{P}_\theta}(Z) = \log \frac{q_\xi(z_0 \mid \Phi)}{p_\theta(z_0)} + \log \frac{d\mathbb{Q}_{\text{SDE}}}{d\mathbb{P}_{\text{SDE}}}(Z \mid z_0).
\end{equation}
The KL divergence is defined as the expectation of this log-ratio under the surrogate measure $\mathbb{E}_{\mathbb{Q}_\xi}$. Taking the expectation of the first term yields the standard KL divergence of the initial state, functioning identically to the regularization term in a Variational Autoencoder (VAE), but interestingly where the initial latent state $z_0$ in conditioned by the entire sequence $\Phi$ and not just $\varphi_0$ :
\begin{equation}
    \mathbb{E}_{\mathbb{Q}_\xi} \left[ \log \frac{q_\xi(z_0 \mid \Phi)}{p_\theta(z_0)} \right] = D_{\text{KL}}(q_\xi(z_0 \mid \Phi) \parallel p_\theta(z_0)).
\end{equation}
For the second term, Girsanov's theorem expresses the path log-ratio as a tractable stochastic integral:
\begin{equation}
    \log \frac{d\mathbb{Q}_{\text{SDE}}}{d\mathbb{P}_{\text{SDE}}}(Z \mid z_0) = \int_0^T u(z_t, t) \cdot dW_t^\mathbb{P} - \frac{1}{2}\int_0^T \|u(z_t, t)\|_2^2 dt.
\end{equation}
Girsanov's theorem also dictates how the driving Brownian motions relate across the two measures. Because $\mathbb{Q}_\xi$ is steered by $u(z_t, t)$, the random increments from the perspective of the prior $\mathbb{P}_\theta$ are biased by the control effort:
\begin{equation}
    dW_t^\mathbb{P} = dW_t^\mathbb{Q} + u(z_t, t)dt.
\end{equation}
Substituting this relationship back into the integral shifts the dynamics entirely to the posterior measure $\mathbb{Q}_\xi$:
\begin{equation}
    \log \frac{d\mathbb{Q}_{\text{SDE}}}{d\mathbb{P}_{\text{SDE}}}(Z \mid z_0) = \int_0^T u(z_t, t) \cdot dW_t^\mathbb{Q} + \frac{1}{2}\int_0^T \|u(z_t, t)\|_2^2 dt.
\end{equation}
Finally, we apply the expectation $\mathbb{E}_{\mathbb{Q}_\xi}$ to evaluate the KL divergence. The first term is an Itô integral with respect to the Brownian motion $dW_t^\mathbb{Q}$. Under standard regularity conditions, this integral is a martingale and its expectation is strictly zero. Thus, the infinite-dimensional KL divergence elegantly collapses to a finite integral over the squared control effort:
\begin{equation}
    D_{\text{KL}}(\mathbb{Q}_{\text{SDE}} \parallel \mathbb{P}_{\text{SDE}}) = \frac{1}{2} \mathbb{E}_{\mathbb{Q}_\xi} \left[ \int_0^T \|u(z_t, t)\|_2^2 dt \right].
\end{equation}
This penalty can be expressed purely in terms of the system's drift and diffusion functions with equation \ref{eq:control_var}:
\begin{equation}
    D_{\text{KL}}(\mathbb{Q}_{\text{SDE}} \parallel \mathbb{P}_{\text{SDE}}) = \frac{1}{2} \mathbb{E}_{\mathbb{Q}_\xi} \left[ \int_0^T \big\| g_\theta^{-1}(z_t, t) \big( h_\xi(z_t, t, \Phi) - f_\theta(z_t, t) \big) \big\|_2^2 dt \right].
\end{equation}
For notational convenience, we define this scaled drift discrepancy as the residual vector $r_{\theta, \xi}(z_t, t) = g_\theta^{-1}(z_t, t) \big( h_\xi(z_t, t, \Phi) - f_\theta(z_t, t) \big)$. Recombining this with the initial state divergence yields the complete KL divergence between the surrogate and prior measures:
\begin{equation}
    D_{\text{KL}}(\mathbb{Q}_\xi \parallel \mathbb{P}_\theta) = D_{\text{KL}}(q_\xi(z_0 \mid \Phi) \parallel p_\theta(z_0)) + \frac{1}{2} \mathbb{E}_{\mathbb{Q}_\xi} \left[ \int_0^T \|r_{\theta, \xi}(z_t, t)\|_2^2 dt \right].
\end{equation}
\end{document}